# Electron-beam Writing of Spectrally Uniform Green Single-photon Emitters in Hexagonal Boron Nitride

## Author Information

### Authors

Qingsong Tao[1], Fuyi Zhou[2], Zhijie Li[3], Yihao Yan[3], Shuangyue Li[1], Yuelan Gao[1], Zijing Wu[1], Yizhou Liu[1], Tao Liang[1], Shuai Yuan[2], Dakun Wu[1], Hongzhi Zhou[1], Qi Zhang[5*], Zhenyi Ni[2*], Chunlei Yu[1,4], Pan Wang[6], Fei Yu[1,4*], Lili Hu[1,4], and Ning Zhou[1*]

### Affiliations

[1] School of Physics and Optoelectronic Engineering, Hangzhou Institute for Advanced Study, University of Chinese Academy of Sciences, Hangzhou 310024, China.

[2] State Key Laboratory of Silicon and Advanced Semiconductor Materials, School of Materials Science and Engineering, Zhejiang University, Hangzhou 310027, China.

[3] Anhui Province Key Laboratory of Scientific Instrument Development and Application, University of Science and Technology of China, Hefei 230026, China.

[4] Shanghai Institute of Optics and Fine Mechanics, Chinese Academy of Sciences, Shanghai 201800, China.

[5] Institute of Quantum Sensing, Institute of Fundamental and Transdisciplinary Research, School of Physics, Zhejiang Key Laboratory of R&D and Application of Cutting-Edge Scientific Instruments, State Key Laboratory of Ocean Sensing, Zhejiang University, Hangzhou 310027, China.

[6] New Cornerstone Science Laboratory, State Key Laboratory of Extreme Photonics and Instrumentation, College of Optical Science and Engineering, Zhejiang University, Hangzhou 310027, China.

[*]Corresponding author Email: qzquantum@zju.edu.cn; zyni@zju.edu.cn; yufei@siom.ac.cn; ningzhou@ucas.ac.cn.

## Abstract

Scalable quantum photonic technologies require single-photon emitters whose positions and emission energies can be engineered simultaneously. Hexagonal boron nitride (hBN) is an attractive room-temperature host, but deterministic creation of spectrally reproducible emitters remains challenging. Here, we use a standard scanning electron microscope as a direct-writing tool to activate bright green single-photon emitters in hBN at predefined sites, without ion implantation or post-fabrication thermal annealing. The written emitters exhibit reproducible zero-phonon-line emission centered near 536 nm, room-temperature antibunching with $g^{(2)}(0)$ as low as 0.08, high brightness, strong linear polarization, and stable emission. Thickness-dependent activation, stacking experiments, cathodoluminescence spectroscopy, and first-principles calculations support a carbon-related defect complex as the most plausible origin of the emission. As a proof of nanophotonic compatibility, we further activate emitters in a nanoparticle-on-mirror plasmonic nanocavity and observe photoluminescence enhancement accompanied by shortened emission lifetimes. These results establish electron-beam direct writing as a practical route to site-selective, spectrally uniform green quantum emitters in hBN, offering a promising basis for integrated room-temperature quantum photonic architectures.

## Introduction

Solid-state single-photon emitters (SPEs) are key quantum light sources for quantum communication, sensing, and integrated photonic information processing[1-3]. For scalable photonic quantum technologies, the central requirement is not to merely generate SPEs, but to achieve arrays in which the location, emission energy and brightness of emitters are reliably engineered. Two-dimensional (2D) hosts are particularly attractive because they offer atomically thin integration, straightforward coupling to nanophotonic structures, and compatibility with heterogeneous device platforms[4-7]. Among these 2D materials, hexagonal boron nitride (hBN) supports bright room-temperature quantum emission and accommodates a rich set of optically active defects within its wide bandgap (~6 eV)[8, 9], with emission spanning from the UV to near-infrared spectrum[10-17]. To date, various methods have been developed to create SPEs within hBN, such as strain engineering[13, 18], ion implantation[19], nanoindentation[20-22], femtosecond laser processing[17, 23], and focused ion beam (FIB) milling[24-27].

Despite rapid progress, a persistent limitation of hBN emitters is the difficulty of achieving spatial and spectral determinism in the same fabrication process. Many approaches can generate emitters with high yield; however, their zero-phonon lines (ZPLs) typically span broad spectral ranges, with most reported emission lying in the yellow-to-red spectral region, and the emitter locations remain stochastic at the nanoscale (Table S1). In contrast, green-emitting hBN SPEs are desirable for visible quantum photonics and for efficient detection with silicon-based single-photon detectors, yet they remain scarce compared with other spectral families[28]. In addition, high-temperature annealing is often required as a pre- or post-processing step to activate SPEs, which complicates integration with pre-patterned nanophotonic structures or heterogeneous devices. A practical route toward scalable hBN quantum photonics therefore requires a direct-write method that combines site selectivity, spectral reproducibility, high brightness, and minimal post-processing.

Electron-beam writing (EBW) has emerged as an effective mask-free approach for SPE fabrication, enabling high spatial resolution and precise positional control[15, 29-32]. EBW has been demonstrated to generate SPEs with specific spectral families (e.g., ZPLs near ~436 nm)[15, 30, 31] or in the yellow range (~573–577 nm)[29, 32]. These advances highlight the potential of EBW as a promising route for deterministic defect activation, yet underscore an open question: can EBW be harnessed as a defect-engineering tool to target new spectral windows without compromising emission stability?

Here, we demonstrate direct EBW of bright green SPEs in hBN using a standard scanning electron microscope (SEM). Using a single annealing-free EBW step without complex defect-engineering pre- or post-processing, we activate emitters at predefined sites, yielding reproducible ZPL emission centered near 536 nm, high room-temperature single-photon purity, strong linear polarization and photostability. Through thickness-dependent activation experiments, deterministic stacking, cathodoluminescence (CL) spectroscopy, and first-principles calculations, we show that the formation of these emitters is governed by a carbon-reservoir-mediated thickness threshold, supporting a carbon-associated defect complex, $V_BC_N^-$, as the plausible microscopic origin. Finally, we demonstrate compatibility with plasmonic nanophotonic architectures by activating emitters in a nanoparticle-on-mirror (NPoM) plasmonic nanocavity, where photoluminescence (PL) enhancement is accompanied by shortened emission lifetimes. Our results fill the gap in the stable creation of highly reproducible green SPEs in hBN, expanding the application prospects of hBN in solid-state SPEs. This work provides a practical direct-write strategy for engineering site-selective, spectrally uniform SPEs in hBN, laying the groundwork for integrated quantum photonic devices based on two-dimensional materials.

## Results

### Direct EBW of site-selective and spectrally uniform green emitters in hBN

Figure 1a schematically illustrates the direct EBW strategy used to activate localized emitters in exfoliated hBN. Mechanically exfoliated hBN flakes were transferred onto $SiO_2$/Si substrates, and predefined arrays of nanoscale exposure sites were patterned using a standard SEM operated at a 10 kV acceleration voltage, 10 µA emission current, and an 8 mm working distance, with an exposure time of 10 s over an area of approximately 150 × 100 $nm^2$ (see Methods). A representative optical micrograph and the corresponding SEM image of the written region are shown in Fig. 1b. The exposed sites in the green-boxed region are spatially registered with the designed spot-array pattern, with four written spots resolved in each orange dashed box. Atomic force microscopy (AFM) confirmed a representative hBN flake thickness of ~120 nm (Fig. 1c). PL mapping under 532-nm continuous-wave (CW) laser excitation reveals localized emission from the EBW-patterned sites (Fig. 1d). The bright PL spots spatially coincide with the designed irradiation array in the green-boxed region of Fig. 1b, indicating that EBW locally activates optically addressable defects. This positional control allows for the creation of designed patterns, exemplified by the bright "UCAS" (University of Chinese Academy of Sciences) logo (Fig. S2). To investigate the spectral uniformity of the written emitters, PL spectra were acquired from four representative sites (marked by orange dashed circles in Fig. 1d) under 488-nm CW laser excitation. As presented in Fig. 1e, all emitters display nearly identical spectral features: a prominent, narrow ZPL at 536 nm accompanied by broad phonon sidebands (PSBs) at approximately 550 nm (1st PSB) and 574 nm (2nd PSB). This reproducible green emission suggests preferential activation of a common green-emitting defect center, rather than stochastic generation of spectrally unrelated luminescent centers.

Notably, the PL spectra also capture the intrinsic $E_{2g}$ Raman mode of the hBN host (denoted by an asterisk in Fig. 1e)[33]. We next used high-resolution Raman spectroscopy to evaluate the lattice structural modification induced by EBW (Fig. 1f). Compared to the pristine region, the irradiated area exhibits a moderate broadening of the hBN $E_{2g}$ mode, with the full width at half maximum (FWHM) increasing from 12.85 $cm^{-1}$ to 15.30 $cm^{-1}$, along with a slight peak redshift from 1366.23 $cm^{-1}$ to 1366.10 $cm^{-1}$. This peak broadening is consistent with increased local lattice disorder and defect density introduced by the electron beam and the observed redshift could potentially arise from local strain (~ 0.046%) or lattice relaxation induced during defect formation[11, 34-36]. Together, the PL and Raman measurements indicate that EBW locally modifies the hBN lattice sufficiently to activate green-emitting defect centers while preserving the crystalline host framework.

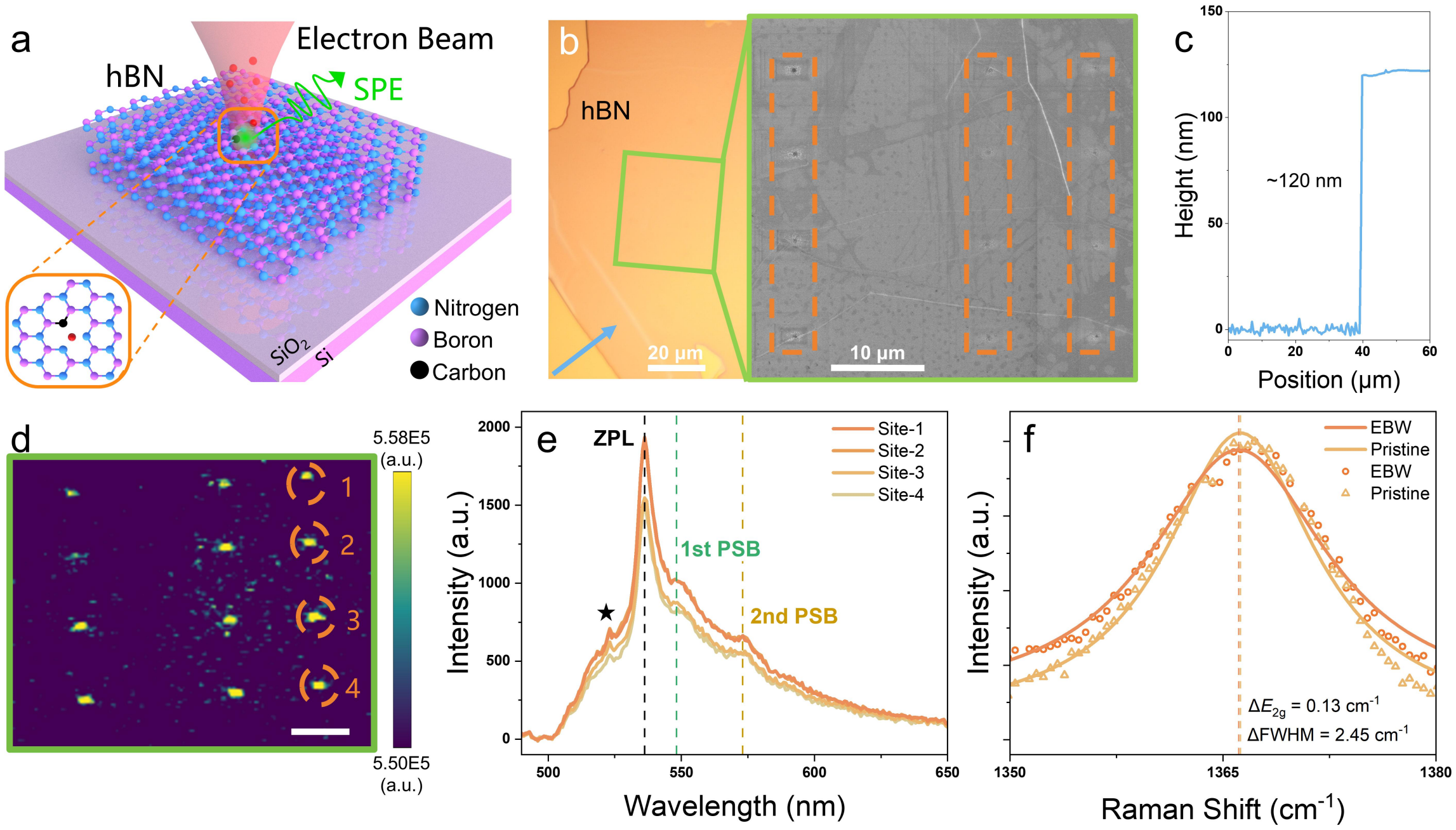

**Fig. 1. Direct EBW of site-selective and spectrally uniform green emitters in hBN.** (**a**) Schematic of localized SPEs activation in an hBN flake by focused electron-beam exposure. (**b**) Optical microscope image of the hBN flake and SEM image of the patterned region. The green box marks the EBW region, and the orange dashed boxes highlight patterned spots. (**c**) AFM height profile along the blue arrow in (b). (**d**) PL map of the EBW-patterned region under 532-nm excitation. Scale bar, 10 µm. (**e**) Room-temperature PL spectra of four written emitters marked by orange circles in (d), showing reproducible ZPL emission near 536 nm and accompanying PSBs. The asterisk denotes the hBN Raman peak. (**f**) Raman spectra acquired from pristine and EBW-exposed regions, showing beam-induced broadening of the hBN $E_{2g}$ mode. Open symbols represent experimental data, and solid lines represent Lorentzian fits.

## Photophysical properties and reproducibility of EBW green SPEs

Figure 2a shows the PL spectra from the same emitter under 488-nm and 532-nm CW laser excitation. Both excitation wavelengths produce the same emission with a sharp ZPL around 536 nm and two broader PSBs near 550 nm (1st PSB) and 574 nm (2nd PSB), indicating that the emission originates from well-defined energy levels intrinsic to the generated defect center. The energy separations between the ZPL and 1st PSB (~58 meV) and 2nd (~153 meV) are consistent with the acoustic phonon modes (~45–60 meV) and the high-energy in-plane optical phonon $E_{2g}$ mode (~160 meV), respectively, indicating effective coupling between the color center and the hBN lattice[37, 38].

The single-photon character of the EBW emitters was verified using a fiber-based Hanbury Brown & Twiss (HBT) setup (Fig. S1). Under 488-nm CW laser excitation, the representative emitter exhibits a clear antibunching dip at zero delay time with a fitted value of $g^{(2)}(0) = 0.08$, well below the classical threshold of 0.5 (Fig. 2b), demonstrating high single-photon purity at room temperature. We consistently observed single-photon emission across multiple fabricated spots, with $g^{(2)}(0)$ values remaining below 0.5 (Fig. S3). All $g^{(2)}(\tau)$ data were fitted using a three-level model described by[39, 40],

$$g^{(2)}(\tau) = y_0 - A\cdot e^{-|\tau|/\tau_1} + B\cdot e^{-|\tau|/\tau_2} \quad (1)$$

where $\tau_1$ and $\tau_2$ are characteristic time constants related to the excited and metastable states. We then measured the excitation-polarization dependence of the defect emission by rotating the linear polarization of the 488 nm excitation laser. The integrated PL intensity exhibits a strong polarization-dependent response as shown in Fig. 2c, which is well-fitted by the function[41],

$$I_\theta = I_{off} + A \cdot \cos^2(\theta - \theta_0) \quad (2)$$

where $I_{off}$ is the background intensity and $\theta_0$ is the offset angle of the dipole. The pronounced excitation-polarization dependence reveals a dipole-like optical response with a fixed in-plane orientation, consistent with a well-defined defect transition in hBN. Such polarization anisotropy enables polarization-selective excitation and optical addressing, offering potential advantages for integration with polarization-sensitive nanophotonic architectures[29].

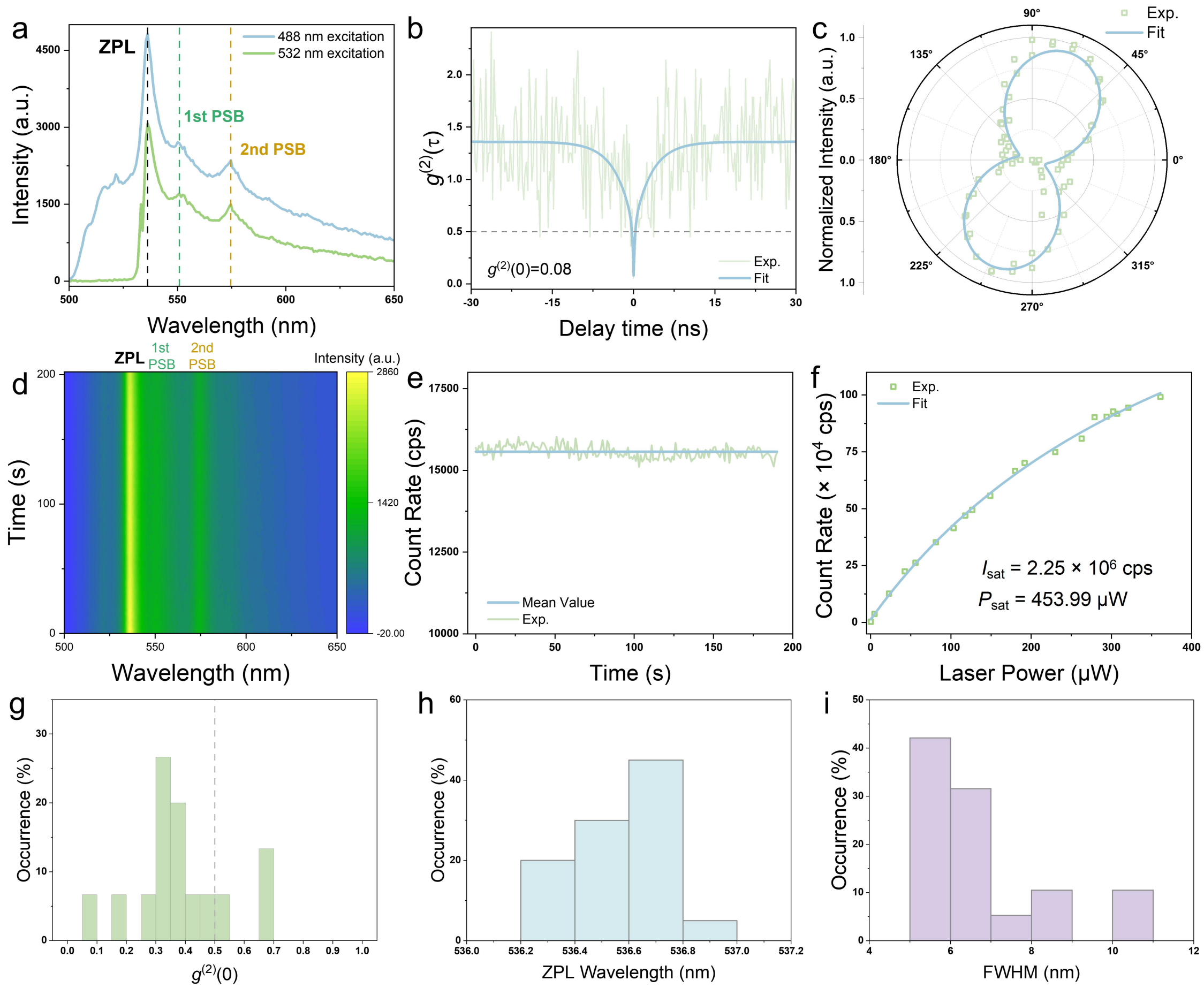

**Fig. 2. Room-temperature photophysical properties and reproducibility of EBW SPEs.** (**a**) PL spectra of a representative emitter under 488-nm (blue) and 532-nm (green) laser excitation. (**b**) Second-order autocorrelation function $g^{(2)}(\tau)$. The fit yields $g^{(2)}(0) = 0.08$. (**c**) Excitation-polarization-dependent PL intensity with fit. (**d**) Time evolution of PL spectra, showing stable ZPL emission over the measurement window. (**e**) Corresponding count rate vs. time, showing stable emission over the measurement window (blue line: mean value). (**f**) Saturation curve with fit (blue line) yielding $I_{sat} = 2.24$ Mcps, $P_{sat} = 454$ µW. (**g**) Histogram of $g^{(2)}(0)$ distribution for characterized EBW SPEs. (**h**) Distribution of the fitted ZPL peak wavelength. (**i**) Distribution of the FWHM of the ZPL.

Photostability under ambient conditions is crucial for practical use of room-temperature SPEs. We therefore monitored the PL spectrum and the emission intensity over time of a representative EBW emitter under 488 nm laser excitation. Over the measurement window, the ZPL position and spectral profile remained stable without noticeable drift (Fig. 2d). The integrated count rate, shown in Fig. 2e, remained constant at around 1.55 × $10^4$ cps, with minor fluctuations, indicating stable optical emission under ambient conditions on the timescale of the measurement. Furthermore, power-dependent PL measurements show a clear saturation behavior (Fig. 2f), which was fitted by the saturation model[32],

$$I = \frac{P \cdot I_{\mathrm{sat}}}{P + P_{\mathrm{sat}}} + I_{\mathrm{d}} \tag{3}$$

where $P$ is the excitation power, $I_{sat}$ is the saturation count rate, $P_{sat}$ is the saturation power, and $I_d$ is the background dark count rate. The fit yields $I_{sat} = 2.24 \times 10^6$ cps and $P_{sat} = 454$ μW. This high detected saturation count rate under our collection and detection configuration highlights the practical brightness of the EBW green SPEs.

To evaluate the reproducibility and yield of the EBW process, we performed $g^{(2)}(\tau)$ measurements on emitters fabricated on a 120-nm-thick hBN flake (raw data shown in Fig. S3). The statistical distribution of $g^{(2)}(0)$ values is presented in Fig. 2g. Among these characterized sites, approximately 80% exhibited $g^{(2)}(0) < 0.5$, indicating that single-photon emission is obtained from a large fraction of the characterized emitters. We further quantified the spectral reproducibility of the EBW emitters by fitting individual ZPLs with Lorentzian profiles. As illustrated in Fig. 2h and i, the ZPL wavelengths are narrowly distributed around a mean of 536.6 nm, with linewidths centered around 5 nm. The narrow ZPL distribution suggests preferential activation of a common green-emitting center with reduced site-to-site spectral inhomogeneity, an important requirement for constructing spectrally uniform emitter arrays.

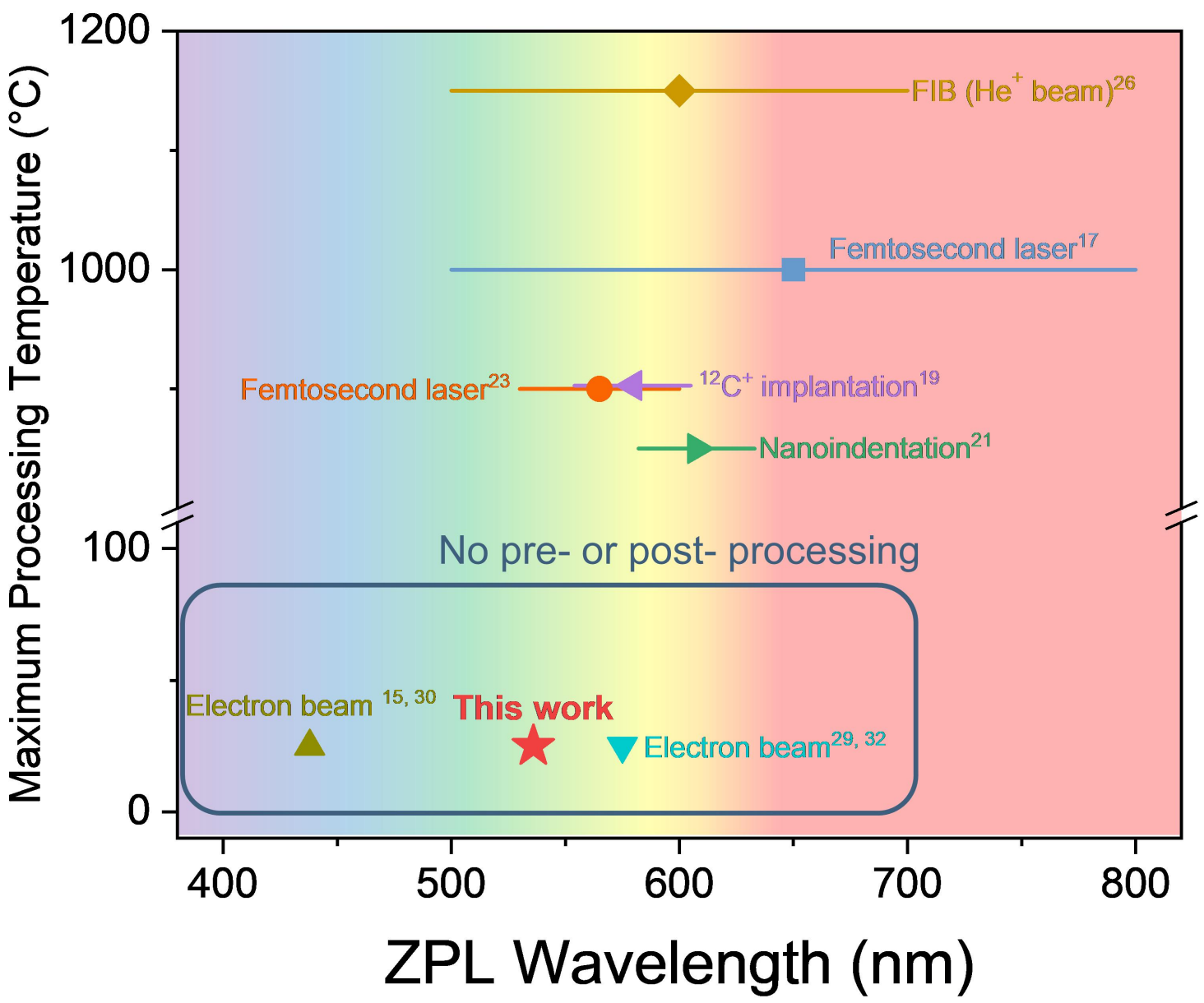

**Fig. 3. Benchmarking of reported room-temperature SPEs in hBN by ZPL wavelength and maximum processing temperature.** Horizontal bars indicate the reported ZPL wavelength coverage of representative hBN SPE fabrication methods, and the vertical axis represents the maximum processing temperature required during emitter generation, including pre- or post-processing steps.

Benchmarking our results against previously reported room-temperature SPEs in hBN highlights the combined advantage of the EBW green emitters in spectral reproducibility and low thermal budget (Figure 3 and Tables S1 and S2). As shown in Fig. 3, many reported fabrication routes either require high-temperature pre- or post-processing or produce emitters over broad ZPL wavelength ranges. In contrast, our EBW approach generates green emitters with a reproducible ZPL near 536 nm under room-temperature processing conditions, without ion implantation or thermal annealing. Together with site-selective activation, high detected saturation count rate, and room-temperature single-photon purity, these features render EBW a scalable tool for deterministic defect engineering in hBN.

**Thickness- and carbon-dependent activation of EBW green emitters**

The formation of the characteristic 536-nm emission shows a clear thickness dependence under otherwise identical EBW conditions. As shown in Figure 4a, hBN flakes thinner than ~20 nm did not show the characteristic 536-nm ZPLs and instead exhibited a broad emission band typical of amorphous carbon defects[33]. Conversely, hBN flakes with thicknesses above 20 nm reproducibly generated the ~536 nm SPEs over a broad thicknesses range from 20 nm to 200 nm (Fig. S4a). To decouple this dimensional constraint from intrinsic material variations, we examined a deterministic stacking structure, schematically illustrated in Fig. 4b, formed by locally stacking three ~15-nm hBN layers to reach a total thickness of ~45 nm. After EBW, the isolated ~15-nm region did not yield detectable SPEs, whereas the locally stacked ~45-nm region showed the characteristic 536-nm emission (Fig. 4c). Monte Carlo electron trajectory simulations further reveal that the incident electrons penetrate through both the 15- and 45-nm hBN layers and reach the underlying $SiO_2$ substrate (Fig. S5), indicating that the absence of emitters in ultrathin flakes is unlikely to result from insufficient electron penetration. These results identify hBN thickness as a key control parameter for EBW-induced activation of the 536-nm emitters.

Furthermore, the essential role of specific intrinsic impurities is unraveled by comparing commercial hBN (HQ Graphene) with high-pressure high-temperature (HPHT) synthesized hBN. Due to intrinsic synthesis imperfections, commercial hBN (HQ graphene) naturally contains a certain concentration of carbon impurities, resulting in a total carbon content that scales with flake thickness. HPHT-grown hBN exhibits higher crystal quality and purity than the commercial hBN crystal[30, 42]. Under the same EBW conditions, HPHT-grown hBN flakes with varying thicknesses failed to generate any SPEs (Fig. S4b). This material dependence suggests that intrinsic carbon impurities are required for efficient activation of the green emitters. Carbon-related defects have been identified as common sources of quantum emission in hBN[43, 44]. We systematically excluded adventitious surface carbon through oxygen plasma treatments. Generating SPEs on pre-cleaned HQ Graphene hBN (Fig. S4c) suggested the internal origin of the carbon. Conversely, post-cleaning samples with already formed SPEs led to a drastic reduction in their original fluorescence (Fig. S4d). The reduced ZPL-to-background ratio after plasma treatment further suggests that the optical response is sensitive to plasma-induced changes in the local defect or surface environment[45, 46]. Finally, a sandwiched structure consisting of a ~15-nm thin HQ Graphene hBN layer enclosed by two ~15-nm thin HPHT hBN layers (Fig. S4e) failed to generate the target emission despite satisfying the total thickness requirement (Fig. S4f). Together, these control experiments support a model in which a sufficient internal carbon-related precursor reservoir, rather than thickness alone, is required for EBW-induced formation of the 536-nm emitters, while macroscopic thickness concurrently offers a larger interaction volume for stable defect stabilization[31].

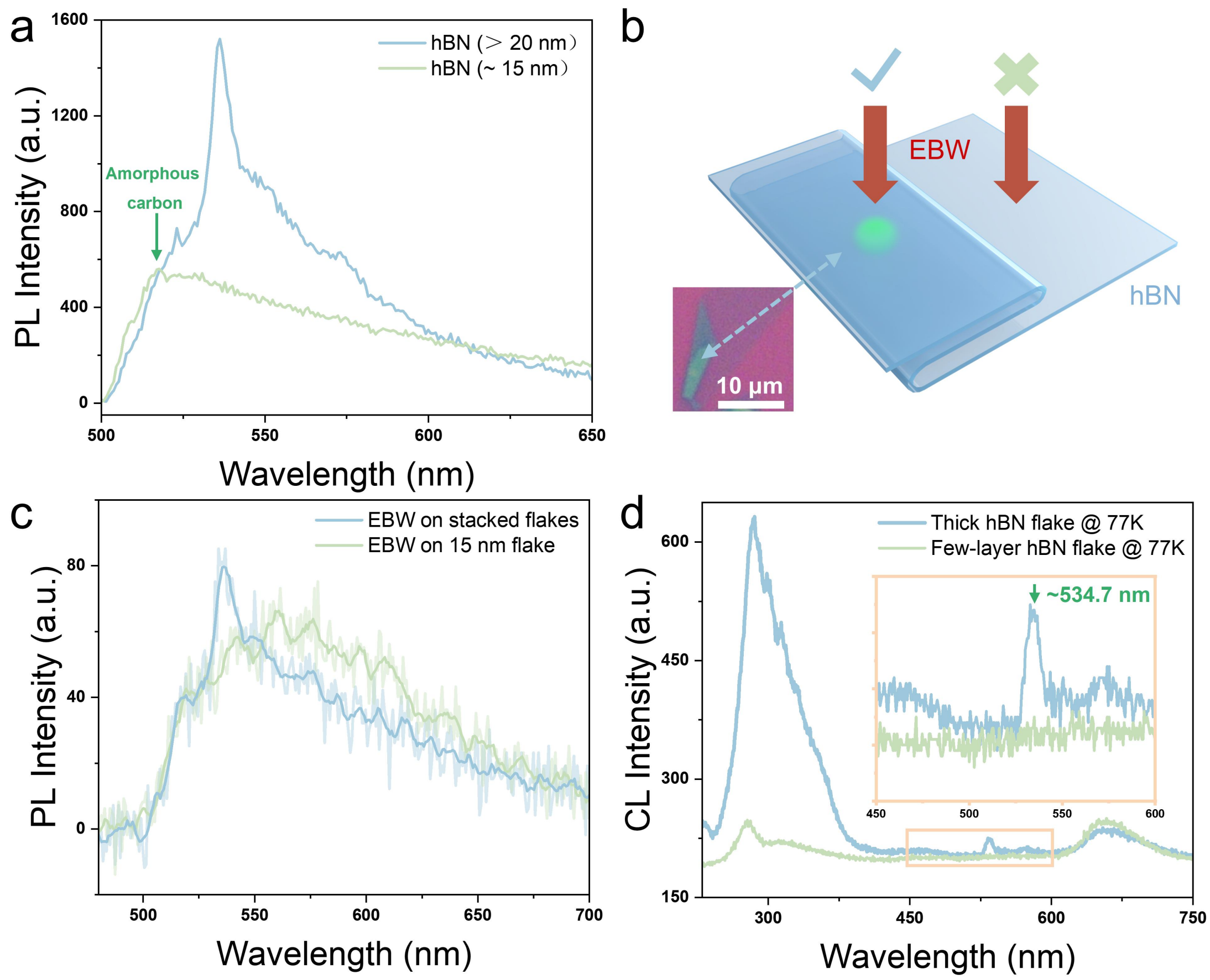

**Fig. 4. Thickness- and carbon-dependent activation of EBW-induced SPEs in hBN.** (**a**) Room-temperature PL spectra of hBN flakes with different thicknesses after EBW. The thick flake (>20 nm, blue) shows a distinct emission peak, whereas the thin sample (<20 nm, green) is dominated by broad emission. (b) Schematic and optical image of the three-layer stacked sample. (**c**) Comparison of PL spectra of stacked hBN flakes with an accumulated thickness of ~45 nm (blue) and an isolated 15 nm flake (green) after EBW.- (**d**) CL spectra of sub-threshold (green) and above-threshold (blue) hBN flakes at 77 K. The inset shows a magnified view of the 450–600 nm region, highlighting the emission of the defect state at 534.7 nm in the thick flake.

The activation process also exhibits a clear dose dependence. As the exposure time increased, the 536-nm ZPL intensity first increased and reached an optimum around 30 s, followed by a gradual decrease under prolonged irradiation (Fig. S6a). This behavior suggests a competition between defect activation and beam-induced damage or quenching[31]. Remarkably, spectrally consistent green emitters could be generated using both cold-field-emission and thermal-field-emission SEM systems (Fig. S6b, c). Taken together, these observations illuminate the role of the electron beam: the EBW process delivers localized energy to trigger the structural rearrangement and activation of pre-existing carbon-related precursors within the hBN lattice.

To further probe defect-related optical states, we performed CL spectroscopy. Both sub-threshold thin and above-threshold thick hBN flakes display a prominent emission peak near 300 nm, consistent with the presence of background carbon-impurity-related defect states[13, 42]. Notably, a sharp emission line at 534.7 nm, close to the ZPL of the EBW green SPEs, emerges only in the thick hBN flakes (Fig. 4d). Room-temperature CL comparison between thick HQ Graphene and HPHT hBN further shows that HQ Graphene hBN exhibits a stronger carbon-related emission near 300 nm, accompanied by a distinct peak around 534 nm, whereas HPHT hBN lacks these signatures (Fig. S7). These CL results are consistent with the thickness- and material-dependent PL observations and further support the involvement of carbon-related precursors in the formation of the 536-nm emitters.

**Candidate carbon-associated vacancy defect modeling for the 536-nm emitter**

Motivated by the carbon-related activation behavior observed experimentally and previous theoretical reports on carbon-associated hBN emitters[32], we examined three candidate carbon–vacancy configurations, $V_NC_N$, $V_NC_B^+$, and $V_BC_N^-$, whose calculated emission energies are close to the experimentally observed 536-nm ZPL. We first recalculated their electronic structures to evaluate whether they contain transition-relevant defect states within the hBN bandgap (Fig. S8). These electronic-structure calculations provide an initial screening of candidate defect levels rather than fully relaxed ZPL energies. Among the three candidates, $V_BC_N^-$ exhibits the most favorable transition-relevant in-gap defect configuration for the experimentally observed 536-nm emission. The atomic structure of the $V_BC_N^-$ complex is shown in Figure 5a. These electronic-structure calculations provide an initial screening of candidate defect levels, but the extracted defect-level separations should not be interpreted as fully relaxed ZPL energies. We next compared the thermodynamic accessibility of the three candidates by calculating their formation energies as a function of the Fermi level (Fig. 5b). $V_NC_B^+$ is mainly competitive near the valence-band edge, whereas the neutral $V_NC_N$ defect shows a moderate, Fermi-level-independent formation energy. In contrast, $V_BC_N^-$ exhibits the lowest formation energy over most of the physically relevant Fermi-level range. Together with its emission energy close to the experimental 536-nm ZPL, this formation-energy trend identifies $V_BC_N^-$ as the most relevant candidate among the three examined configurations.

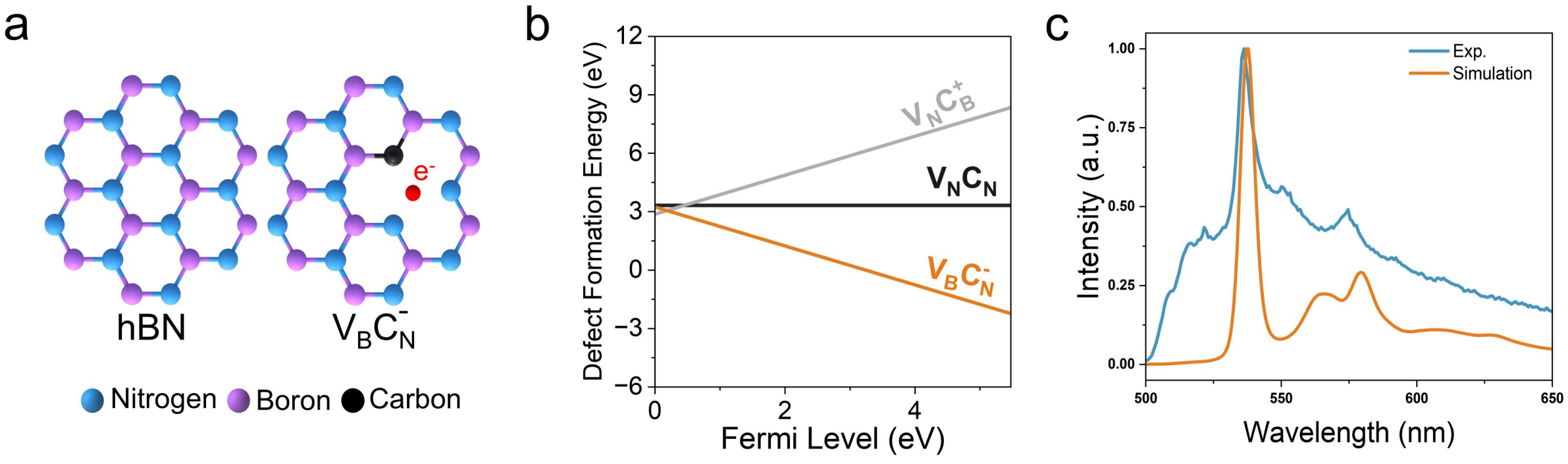

**Fig. 5. Candidate carbon-associated vacancy-complex defect modeling for the 536-nm emitter.** (a) Atomic structure of the pristine hBN lattice (left) and the modeled negatively charged boron vacancy-carbon antisite defect (right). (**b**) Calculated formation energies of three candidate carbon-associated defects, $V_NC_N$, $V_NC_B^+$, and $V_BC_N^-$, plotted as a function of the Fermi level. The vertical dashed line marks the conduction-band minimum $E_{CBM}$. (c) Comparison between the experimental PL spectrum (blue) and the vibrationally resolved TD-DFT spectrum (orange) calculated for $V_BC_N^-$, showing qualitative agreement in the ZPL position and vibronic sideband profile.

To evaluate whether this defect candidate can account for the observed vibronic structure, we calculated the vibrationally resolved emission spectrum using time-dependent density functional theory (TD-DFT) within a finite cluster model. Following geometry optimizations, the emission spectrum was computed using the Adiabatic Hessian (AH) model, which accounts for geometry relaxation between the ground and excited states[47]. Furthermore, both Franck-Condon (FC) and Herzberg-Teller (HT) contributions were evaluated, and the Tamm-Dancoff Approximation (TDA) was used to improve numerical stability in the excited-state calculations[48, 49]. As shown in Fig. 5c, the calculated spectrum exhibits qualitative agreement with the main experimental features, including the ZPL position and the PSB profile. This qualitative spectral agreement, together with the carbon-dependent control experiments and CL data, supports $V_BC_N^-$ as the plausible microscopic candidate for the 536-nm emitter. We also performed ODMR measurements to search for optically addressable spin transitions. No resolvable ODMR contrast was observed over 600-5000 MHz at zero field or under a 50 mT magnetic field under the present measurement conditions (Fig. S9). Therefore, we do not assign the spin multiplicity or a spin-selective optical cycle in this work.

**Proof-of-concept integration with NPoM plasmonic structure**

One important advantage of layered hBN is its facile integration with nanophotonic architectures. Coupling quantum emitters with nanoparticle on mirror (NPoM) plasmonic architectures has shown tremendous potential for tailoring optical, electrical, and spin properties at the nanoscale[24, 50-54]. As activation of the 536-nm emitters requires an hBN thickness of approximately 20 nm, the resulting hBN spacer in NPoM is thicker than the few-nanometer gaps typically used to achieve optimum PL enhancement[55]. We therefore used the NPoM geometry as a proof-of-concept platform to assess whether EBW emitters can be incorporated into a plasmonic nanophotonic environment. We constructed an NPoM architecture consisting of a ~20 nm hBN spacer between a 100 nm Au nanosphere (NS) and an Au microflake (Figure 6a). By directing the electron beam to the Au NS region, we activated ~536-nm emission from the NPoM-associated hBN region. Correlative bright-field optical microscopy and SEM confirm the location and morphology of the NPoM structure (Fig. 6b). Single-particle dark-field scattering measurements showed a dominant plasmonic resonance near 720 nm, together with weaker visible-band response extending across the excitation and emission spectral region of the green emitters. The measured scattering response was qualitatively captured by electromagnetic simulations (Fig. 6c). Notably, the scattering response at 532 nm is stronger than that at 488 nm, suggesting that the NPoM structure can provide a more pronounced plasmonic antenna response under 532-nm excitation. We then compared the PL intensity from the NPoM-associated written region with that from nearby EBW-written emitters in bare hBN regions on the same Au microflake. In the absence of the Au nanosphere, the reference emitters show stronger PL under 488-nm excitation than under 532-nm excitation at the same excitation power, indicating that the 536-nm defect is intrinsically more efficiently excited at 488 nm. After integration with the NPoM structure, the ZPL emission is enhanced relative to the corresponding bare-hBN reference for both excitation wavelengths. The ZPL enhancement factor, defined as the integrated ZPL intensity ratio between the NPoM-associated region and a nearby nanoparticle-free hBN region on the same Au film, is ~2.15 under 488-nm excitation and ~4.53 under 532-nm excitation (Fig. 6d). HBT measurements of the NPoM-associated emitters show clear antibunching with $g^2(0) = 0.46 < 0.5$, confirming that the single-photon emission characteristics are preserved after integration with the plasmonic structure (Fig. 6e). Time-resolved PL measurements further show that the emission lifetime is shortened from ~1.16 ns for the reference SPEs to ~0.25 ns for the NPoM-associated emitters (Fig. 6f), indicating modified decay dynamics in the plasmonic environment. Previous CL studies have shown that focused electron beams can excite gap-associated plasmonic modes in NPoM structures[56, 57]. Together with the lifetime shortening observed for the NPoM-associated emitters, these results are consistent with the possibility that targeted EBW exposure at the Au NS couples energy into the NPoM gap region and promotes defect activation in the associated hBN spacer. However, this evidence remains indirect, and we do not exclude contributions from emitters formed near, rather than strictly inside, the nanogap. Moreover, we do not exclude contributions from emitters formed near, rather than strictly inside the nanogap. Although the absolute PL intensity under 532-nm excitation remains lower than that obtained under 488-nm excitation, the relative NPoM-associated enhancement is stronger at 532 nm, consistent with the stronger dark-field scattering response of the NPoM structure near this wavelength. These results indicate that the present NPoM design provides an initial demonstration of compatibility between EBW hBN emitters and plasmonic nanophotonic structures, together with modest excitation-dependent PL enhancement and lifetime modification. Further optimization of the hBN spacer thickness, cavity geometry, and emitter–field spatial overlap will be needed to achieve stronger Purcell enhancement or efficient ZPL-rate enhancement.

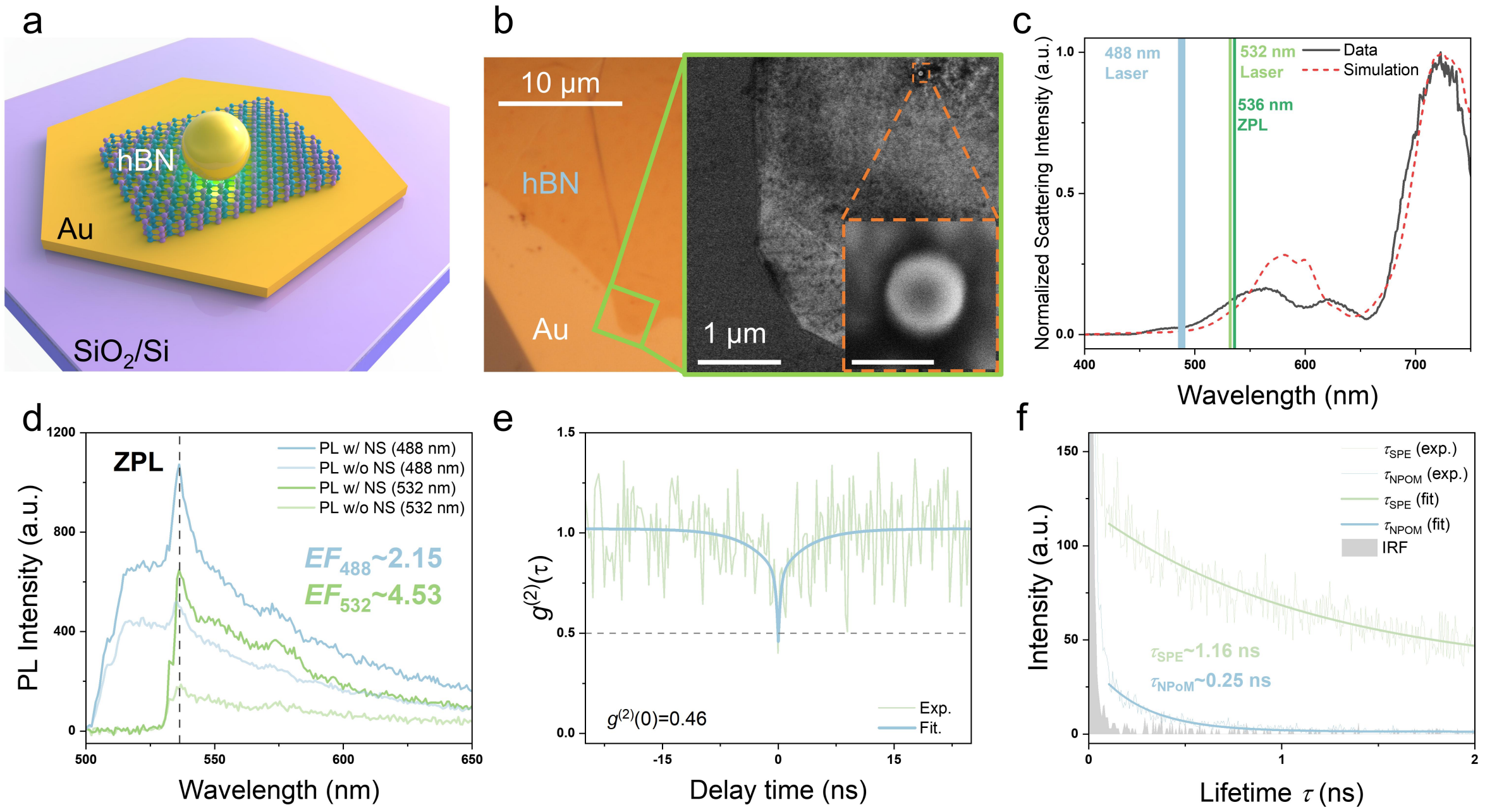

**Fig. 6. Proof-of-concept integration of EBW-engineered green SPEs with an NPoM plasmonic structure.** (a) Schematic illustration of the hybrid NPoM architecture. (b) Correlative optical bright-field image (left) and SEM (right) of the NPoM structures. The inset shows a magnified SEM view of an individual 100-nm Au NS (indicated by the dashed orange box, scale bar: 100 nm). (c) Experimental (solid grey line) and simulated (dashed red line) dark-field scattering spectra of a single NPoM cavity with a 20-nm hBN spacer. The vertical lines indicate the 488-nm and 532-nm excitation wavelengths and the targeted SPE ZPL (~536 nm). (d) PL spectra comparing the NPoM-associated written region with nearby EBW reference regions in bare hBN on the same Au microflake under 488-nm (blue) and 532-nm (green) laser excitation. (e) $g^{(2)}(\tau)$ of an NPoM-associated emitter, showing antibunching with ($g^{(2)}(0) = 0.46 < 0.5$). (f) Time-resolved PL decay curves of reference SPEs and NPoM-associated emitters. The instrument response function (IRF) is also shown.

Thus, although the absolute PL intensity under 532-nm excitation remains lower than that obtained under 488-nm excitation, the relative NPoM-induced enhancement is stronger at 532 nm, consistent with the stronger dark-field scattering response of the NPoM structure near this wavelength. These results indicate that the present NPoM design provides an initial demonstration of compatibility between EBW hBN emitters and plasmonic nanophotonic structures, together with modest excitation-dependent PL enhancement. Further optimization of the hBN spacer thickness, cavity geometry, and emitter-field spatial overlap will be needed to achieve stronger Purcell enhancement or efficient ZPL-rate enhancement.

## Discussion

This work establishes EBW as an annealing-free strategy for activating site-registered green SPEs in hBN. Simply using a standard SEM, we locally activated emitters at predefined sites without ion implantation or post-fabrication thermal annealing. The written emitters exhibit reproducible ZPL emission near 536 nm, room-temperature antibunching, pronounced dipolar optical response, stable emission, and high brightness. Together, these properties identify the EBW centers as practically bright green hBN SPEs and highlight SEM-based direct writing as an accessible route for defect activation in layered quantum materials.

Beyond site-registered activation, our results suggest a precursor-reservoir model for the formation of the 536-nm emitters. The absence of the characteristic ZPL in sub-20-nm flakes, its recovery in stacked hBN, the absence of target emission in high-purity HPHT hBN, and the CL signatures of commercial HQ Graphene hBN collectively support the carbon-related precursors in the activation process. The electron beam does not merely introduce lattice damage; rather, it likely drives bond rearrangement, charge-state modification, or defect stabilization of pre-existing carbon-related precursors within an effective hBN interaction volume. Guided by the experimentally observed carbon-related activation behavior and previously reported carbon-associated hBN defect libraries, the vibronically resolved simulations further support a carbon-vacancy defect complex $V_BC_N^-$ as a plausible microscopic candidate for the 536-nm emission. As a proof of nanophotonic compatibility, we integrated the EBW-written emitters with an NPoM plasmonic architecture, where excitation-wavelength-dependent PL enhancement and shortened emission lifetimes were observed.

The reproducible green emission near 536 nm lies in a spectral region compatible with efficient silicon-based single-photon detection and common visible photonic components. More broadly, SEM-based activation of spectrally reproducible emitters provides a practical materials-processing route toward site-registered hBN quantum-emitter arrays. When combined with controlled carbon-precursor engineering, wafer-compatible hBN platforms, and resonantly designed photonic or plasmonic architectures, this strategy could transform stochastic defect activation in hBN into a manufacturable framework for room-temperature integrated quantum photonics.

## Methods

**hBN sample preparation and deterministic transfer**

Bulk hBN crystals from HQ Graphene and HPHT-grown hBN crystals from Taizhou Sunano Energy Co., Ltd. were used as source materials. hBN flakes were first mechanically exfoliated from bulk hBN crystal using standard Scotch tape and then transferred onto a polydimethylsiloxane (PDMS) substrate. Selected flakes were subsequently dry-transferred onto cleaned $SiO_2$/Si substrates using a deterministic transfer system (Metatest E1). The $SiO_2$ thickness was 300 nm. No thermal annealing was applied to the hBN flakes before or after EBW. The hBN flake thickness was determined using AFM (Dimension Icon, Bruker). For thickness-dependent and stacking experiments, the thickness of each individual flake and the final stacked region was verified by AFM before optical characterization.

**EBW of green emitters**

EBW was performed using a field-emission SEM. For the main EBW experiments, a cold-field-emission SEM (SU8010, Hitachi) was operated at an acceleration voltage of 10 kV and a working distance of 8 mm. Predefined rectangular regions of approximately 150 × 100 nm$^2$ were exposed for 10 s at each writing site. The nominal emission current was 10 μA. For comparison between SEM source types, additional EBW experiments were performed using a thermal-field-emission SEM (Zeiss Sigma) at an acceleration voltage of 3 kV with the same nominal exposure time. These experiments were used to test qualitative reproducibility across SEM systems.

**NPoM plasmonic-structure fabrication**

Single-crystal gold flakes with a thickness of 50 μm were synthesized using a wet chemical method[50]. First, a growth solution was prepared by the addition of 90 μL of 0.1 M chloroauric acid (98%, Sigma-Aldrich) aqueous solution into a 10 mL ethylene glycol (99.8%, Sigma-Aldrich) in a 20 mL glass vial. Then, a cleaned $SiO_2$/Si substrate with a slightly tilted angle was immersed into the mixed solution at a slightly tilted angle. The growth solution was heated to 95°C in an oven for 24 hours. Finally, the gold flakes on $SiO_2$/Si substrate were then immersed in acetone and ethanol for ultrasonic cleaning and dried in a nitrogen environment for the next step use.

hBN flakes with a thickness of approximately 20 nm were transferred onto the Au microflakes to serve as dielectric spacers. Au NSs with a diameter of 100 nm (from NanoSeedz Ltd.) were diluted and a microdrop of the solution was dipped onto the hBN/Au substrate. Excess solution was removed using a gentle $N_2$ flow. Isolated Au NSs were identified by optical microscopy and SEM before EBW exposure.

For NPoM-associated emitter activation, the electron beam was directed to the Au-nanosphere region using the same EBW protocol as used for nearby hBN reference regions. Reference emitters were written in hBN regions on the same Au microflake but away from Au nanospheres.

**Photophysical characterization of EBW green SPEs**

Optical microscopy images, PL, and Raman spectra were acquired using a custom-built setup based on an upright optical microscope (LV100ND, Nikon; see Fig. S1) equipped with a CCD camera (DS-Fi3, Nikon). PL spectra were recorded using a fiber-coupled spectrometer (QE Pro, Ocean Optics), while Raman measurements utilized a monochromator (Omni-A3004i, Zolix) coupled with an electron-multiplying CCD (EMCCD; DU-888, Andor). The hBN samples were excited by CW lasers operating at 488 nm (LDF488-20, JCOPTIX) or 532 nm (MSL-FN-532, Changchun New Industries Optoelectronics Tech. Co., Ltd.), focused to a spot area of approximately 1.5 $\mu m^2$ using a 100× objective lens (NA = 0.9, Nikon). The emission signal was separated from the excitation path using appropriate dichroic beam splitters (DMLP505T, Thorlabs; Di03-R532, Semrock) and long-pass filters (DOFE1LP-500, JCOPTIX; LP03-532RE, Semrock), respectively. Additional PL mapping was performed using a commercial confocal Raman imaging system (WITec) at Westlake University under 532-nm CW laser excitation at 5 mW.

Second-order photon correlation measurements, $g^{(2)}(\tau)$, were conducted using a fiber-based HBT configuration. The collected PL signal under 488-nm CW laser excitation at 80 μW was spectrally filtered and then coupled into a 1×2 50:50 single-mode fiber splitter (TN532R5F1, Thorlabs). The two output arms were connected to two avalanche photodiode (APD) detectors (SPCM-CD3808H, Excelitas), and photon correlations were recorded using a time-tagging module (Time Tagger 20, Swabian Instruments). For photostability measurements, the PL spectrum and emission intensity of a representative EBW emitter were monitored over time under CW 488-nm laser excitation at an excitation power of 50 μW.

Time-resolved PL measurements of both reference EBW emitters on Au flakes and NPoM-associated emitters were performed using a 488-nm femtosecond laser generated from an optical parametric amplifier (OPA) as the excitation source. The excitation power was set to 50 μW. The emitted photons were spectrally filtered using a 515-nm long-pass filter and a 550-nm short-pass filter to suppress the excitation laser and reduce contributions from the phonon sidebands, thereby isolating the ZPL-dominated emission. The PL decay signals were recorded using the same fiber-coupled single-photon detection and time-tagging system described above. The IRF was measured separately from the scattered excitation pulse collected on a nearby Au microflake and was used as the temporal response reference for lifetime analysis.

CL spectroscopy was conducted inside a field-emission SEM (FEG-SEM; Quanta 200F, Thermo Fisher Scientific). Light was collected using a high-sensitivity CL detector (Rainbow, Goldenscope Technology) featuring a retractable ellipsoidal mirror and fiber-optic coupling to a high-resolution monochromator/CCD assembly. Low-temperature CL measurements were performed at 77 K.

**ODMR measurements**

ODMR measurements were performed at room temperature using a home-built confocal microscope on individual single-photon emitters in an hBN thin film supported on a gold substrate. Optical excitation was provided by a continuous-wave 532 nm laser, and the photoluminescence was collected within a 550-630 nm spectral window. Microwave excitation was applied through an Ω-shaped coplanar waveguide. Continuous-wave measurements were performed by sweeping the microwave frequency from 600 to 5000 MHz while monitoring the photoluminescence intensity. Measurements were conducted on three emitters under zero applied magnetic field and on 38 spatially distinct emitters within the same hBN film under an out-of-plane magnetic field of 50 mT. The signal stability was quantified as the standard deviation of the normalized photoluminescence intensity at each microwave frequency point across repeated full ODMR frequency sweeps, yielding a typical variation of ~0.5%.

**Thickness, stacking, plasma-treatment, and material-control experiments**

For thickness-dependent activation experiments, hBN flakes with thicknesses ranging from 15 nm to 200 nm were exposed using identical EBW conditions. For deterministic stacking experiments, selected hBN flakes with individual thicknesses of approximately 15 nm were sequentially transferred onto the substrate using the dry-transfer system. The final thickness of the stacked region was verified by AFM. EBW was then performed on both isolated thin-flake regions and locally stacked regions under identical conditions.

Oxygen-plasma treatment was performed either before EBW exposure or after emitter formation. The treatment was conducted in a vacuum plasma cleaner at a chamber pressure of 100 Pa for 30 min using oxygen plasma. Plasma-treated and untreated regions were characterized under identical PL conditions.

For material-control experiments, HQ Graphene hBN and HPHT-grown hBN flakes were exposed under identical EBW and PL measurement conditions. The absence or presence of the characteristic 536-nm ZPL was evaluated using the same spectral criteria.

**Monte Carlo electron-trajectory simulations**

Monte Carlo simulations of electron trajectories were performed using CASINO v2.48. The simulated structure consisted of hBN layers with thicknesses of 15 nm and 45 nm on $SiO_2$/Si substrates. The incident electron energy was set to 10 keV to match the EBW conditions, with 10,000 electron trajectories simulated for each structure, a beam radius of 100 nm, and a specimen tilt angle of 0°. Both electron trajectories and energy-deposition profiles were analyzed. The simulations were used to evaluate whether the electron range exceeds the hBN thickness under the EBW conditions, but not to uniquely determine the microscopic defect-activation mechanism.

**Electronic-structure and vibronically resolved emission-spectrum calculations**

First-principles calculations based on DFT were performed using the Vienna Ab initio Simulation Package (VASP, version 6.4.1)[58]. The projector-augmented wave (PAW) method was employed to describe the ion-electron interactions[59]. The exchange-correlation interactions were modeled using the generalized gradient approximation (GGA) in the Perdew-Burke-Ernzerhof (PBE) formulation[60, 61]. To properly describe the monolayer hBN sheet, a vacuum space of 15 Å was applied along the z-direction to eliminate spurious periodic interactions between adjacent layers. Long-range van der Waals (vdW) interactions were accounted for using the DFT-D3 dispersion correction method proposed by Grimme[62]. The geometric structure was fully optimized during the PBE stage. The Brillouin zone was sampled using a Γ-centered 4×4×1 Monkhorst-Pack k-point grid. The plane-wave kinetic energy cutoff was set to 500 eV. During relaxation, the lattice parameters in the xy plane and all atomic positions were allowed to relax, while the vacuum thickness along the z-direction was kept fixed. The atomic positions were fully relaxed with the lattice constants fixed at experimental values. The convergence criteria for electronic self-consistency and ionic relaxation were set to $10^{-6}$ eV and 0.01 eV/Å, respectively. To overcome the systemic bandgap underestimation of the PBE functional, electronic properties and band structures were investigated using the Heyd-Scuseria-Ernzerhof (HSE06) hybrid functional[63]. The self-consistent field (SCF) calculations under the HSE06 scheme were performed on a denser k-point grid. The band structure was subsequently calculated along the high-symmetry path within VASP. The density of states (DOS) was obtained using the same 8×8×1 grid with the tetrahedron method with Blöchl corrections (ISMEAR = -5).

TD-DFT calculations, including geometry optimizations and vibrational frequency analyses for both the electronic ground and first excited states, were performed using the Gaussian 16 software package. Structural relaxations utilized the hybrid PBE0 functional in combination with the 6-311G(d) basis set[64, 65]. To accurately mimic the rigid macroscopic crystal environment and preserve the structural integrity of the finite cluster model, the saturating hydrogen atoms at the boundaries and their directly bonded heavy atoms were kept frozen during optimizations. Frequency analyses confirmed that the optimized structures correspond to true local minima (no imaginary frequencies) and generated the Hessian matrices required for vibronic coupling calculations. The vibrationally resolved emission spectra were subsequently simulated using the excited state dynamics (ESD) module implemented in the ORCA 6.1 program[66], with necessary input files generated via the Multiwfn code (version 3.8)[48]. All ESD simulations were conducted at a standardized temperature of 298 K. The final theoretical spectra were broadened using Gaussian functions with a FWHM of 150 $cm^{-1}$ to strictly match the experimental spectral resolution.

**Electromagnetic simulations of NPoM scattering**

Electromagnetic simulations of the NPoM structure were performed using the finite-difference time-domain simulations (Ansys Lumerical FDTD). The simulated structure was constructed according to the experimental geometry, comprising a 100-nm-diameter Au nanosphere positioned on a 20-nm-thick hBN spacer layer supported by a 50-nm-thick Au film on a $SiO_2$/Si substrate. The refractive index of hBN was set to 2.13, and the optical constants of Au were taken from Johnson and Christy. A TM-polarized plane wave was used as the excitation source and was incident at an angle of 30° with respect to the surface normal, consistent with the experimental dark-field configuration. Perfectly matched layer boundary conditions were applied in all directions. A refined mesh was applied in the nanogap region to accurately resolve the localized plasmonic field. The extinction spectrum was obtained from the calculated extinction cross section.

## References

1. M. Sutula, I. Christen, E. Bersin, M. P. Walsh, K. C. Chen, J. Mallek, A. Melville, M. Titze, E. S. Bielejec, S. Hamilton, D. Braje, P. B. Dixon, D. R. Englund, Large-scale optical characterization of solid-state quantum emitters. *Nat. Mater.* **22**, 1338-1344 (2023).
2. I. Aharonovich, D. Englund, M. Toth, Solid-state single-photon emitters. *Nat. Photonics* **10**, 631-641 (2016).
3. M. Pompili, S. L. N. Hermans, S. Baier, H. K. C. Beukers, P. C. Humphreys, R. N. Schouten, R. F. L. Vermeulen, M. J. Tiggelman, L. dos Santos Martins, B. Dirkse, S. Wehner, R. Hanson, Realization of a multinode quantum network of remote solid-state qubits. *Science* **372**, 259-264 (2021).
4. W. Wang, L. O. Jones, J. S. Chen, G. C. Schatz, X. Ma, Utilizing ultraviolet photons to generate single-photon emitters in semiconductor monolayers. *ACS Nano* **16**, 21240-21247 (2022).
5. L. Peng, H. Chan, P. Choo, T. W. Odom, S. Sankaranarayanan, X. Ma, Creation of single-photon emitters in $WSe_2$ monolayers using nanometer-sized gold tips. *Nano Lett.* **20**, 5866-5872 (2020).
6. M. Koperski, K. Nogajewski, A. Arora, V. Cherkez, P. Mallet, J. Y. Veuillen, J. Marcus, P. Kossacki, M. Potemski, Single photon emitters in exfoliated $WSe_2$ structures. *Nat. Nanotechnol.* **10**, 503-506 (2015).
7. J. Lee, H. Wang, K. Y. Park, S. Huh, D. Kim, M. Yu, C. Kim, K. S. Thygesen, J. Lee, Room temperature quantum emitters in van der Waals alpha-$MoO_3$. *Nano Lett.* **25**, 1142-1149 (2025).
8. Y. Han, S. Feng, K. Cao, Y. Wang, L. Gao, Z. Xu, Y. Lu, Large elastic deformation and defect tolerance of hexagonal boron nitride monolayers. *Cell Rep. Phys. Sci.* **1**, 100172 (2020).
9. C. Su, E. Janzen, M. He, C. Li, A. Zettl, J. D. Caldwell, J. H. Edgar, I. Aharonovich, Fundamentals and emerging optical applications of hexagonal boron nitride: a tutorial. *Adv. Opt. Photonics* **16**, 229-346 (2024).
10. R. Bourrellier, S. Meuret, A. Tararan, O. Stephan, M. Kociak, L. H. Tizei, A. Zobelli, Bright UV single photon emission at point defects in h-BN. *Nano Lett.* **16**, 4317-4321 (2016).
11. T. W. Tang, R. Ritika, M. Tamtaji, H. Liu, Y. Hu, Z. Liu, P. R. Galligan, M. Xu, J. Shen, J. Wang, J. You, Y. Li, G. Chen, I. Aharonovich, Z. Luo, Structured-defect engineering of hexagonal boron nitride for identified visible single-photon emitters. *ACS Nano* **19**, 8509-8519 (2025).
12. M. Yu, J. Lee, K. Watanabe, T. Taniguchi, J. Lee, Electrically pumped h-BN single-photon emission in van der Waals heterostructure. *ACS Nano* **19**, 504-511 (2025).
13. X. Chen, X. Yue, L. Zhang, X. Xu, F. Liu, M. Feng, Z. Hu, Y. Yan, J. Scheuer, X. Fu, Activated single photon emitters and enhanced deep-level emissions in hexagonal boron nitride strain crystal. *Adv. Funct. Mater.* **34**, 2306128 (2024).
14. M. A. Sakib, B. Triplett, W. Harris, N. Hussain, A. Senichev, M. Momenzadeh, J. Bocanegra, P. Vabishchevich, R. Wu, A. Boltasseva, V. M. Shalaev, M. R. Shcherbakov, Purcell-induced bright single photon emitters in hexagonal boron nitride. *Nano Lett.* **24**, 12390-12397 (2024).
15. C. Fournier, A. Plaud, S. Roux, A. Pierret, M. Rosticher, K. Watanabe, T. Taniguchi, S. Buil, X. Quélin, J. Barjon, J.-P. Hermier, A. Delteil, Position-controlled quantum emitters with reproducible emission wavelength in hexagonal boron nitride. *Nat. Commun.* **12**, 3779 (2021).

16. K. Konthasinghe, C. Chakraborty, N. Mathur, L. Qiu, A. Mukherjee, G. D. Fuchs, A. N. Vamivakas, Rabi oscillations and resonance fluorescence from a single hexagonal boron nitride quantum emitter. *Optica* **6**, 542–548 (2019).
17. X. J. Wang, H. H. Fang, Z. Z. Li, D. Wang, H. B. Sun, Laser manufacturing of spatial resolution approaching quantum limit. *Light Sci. Appl.* **13**, 6 (2024).
18. N. V. Proscia, Z. Shotan, H. Jayakumar, P. Reddy, C. Cohen, M. Dollar, A. Alkauskas, M. Doherty, C. A. Meriles, V. M. Menon, Near-deterministic activation of room-temperature quantum emitters in hexagonal boron nitride. *Optica* **5**, 1128–1134 (2018).
19. D. Zhong, S. Gao, M. Saccone, J. R. Greer, M. Bernardi, S. Nadj-Perge, A. Faraon, Carbon-related quantum emitter in hexagonal boron nitride with homogeneous energy and 3-fold polarization. *Nano Lett.* **24**, 1106-1113 (2024).
20. S. L. Ahmed, L. Harsch, C. Imre, I. Karapatzakis, L. Kussi, J. Resch, M. Schrodin, I. Hausler, T. Wagner, C. T. Koch, C. Surgers, W. Wernsdorfer, Nanoindentation for tailored single-photon emitters in hBN: influence of annealing on defect stability. *ACS Nano* **19**, 36302–36312 (2025).
21. X. Xu, Z. O. Martin, D. Sychev, A. S. Lagutchev, Y. P. Chen, T. Taniguchi, K. Watanabe, V. M. Shalaev, A. Boltasseva, Creating quantum emitters in hexagonal boron nitride deterministically on chip-compatible substrates. *Nano Lett.* **21**, 8182-8189 (2021).
22. M. Luo, J. Ge, P. Huang, Y. Yu, I. C. Seo, K. Lu, H. Sun, J. K. Tan, B. K. Tay, S. Kim, W. Gao, H. Li, D. Nam, Deterministic formation of carbon-functionalized quantum emitters in hexagonal boron nitride. *Nat. Commun.* **16**, 11450 (2025).
23. L. Gan, D. Zhang, R. Zhang, Q. Zhang, H. Sun, Y. Li, C. Z. Ning, Large-scale, high-yield laser fabrication of bright and pure single-photon emitters at room temperature in hexagonal boron nitride. *ACS Nano* **16**, 14254-14261 (2022).
24. N. Mendelson, R. Ritika, M. Kianinia, J. Scott, S. Kim, J. E. Froch, C. Gazzana, M. Westerhausen, L. Xiao, S. S. Mohajerani, S. Strauf, M. Toth, I. Aharonovich, Z. Q. Xu, Coupling spin defects in a layered material to nanoscale plasmonic cavities. *Adv. Mater.* **34**, e2106046 (2023).
25. H. Liang, Y. Chen, C. Yang, K. Watanabe, T. Taniguchi, G. Eda, A. A. Bettiol, High sensitivity spin defects in hBN created by high-energy He beam irradiation. *Adv. Opt. Mater.* **11**, 2201941 (2022).
26. G. L. Liu, X. Y. Wu, P. T. Jing, Z. Cheng, D. Zhan, Y. Bao, J. X. Yan, H. Xu, L. G. Zhang, B. H. Li, K. W. Liu, L. Liu, D. Z. Shen, Single photon emitters in hexagonal boron nitride fabricated by focused helium ion beam. *Adv. Opt. Mater.* **12**, 2302083 (2024).
27. J. Ziegler, R. Klaiss, A. Blaikie, D. Miller, V. R. Horowitz, B. J. Aleman, Deterministic quantum emitter formation in hexagonal boron nitride via controlled edge creation. *Nano Lett.* **19**, 2121-2127 (2019).
28. K. S. McKay, J. Kim, H. H. Hogue, Enhanced Quantum Efficiency of the Visible Light Photon Counter in the Ultraviolet Wavelengths. *Opt. Express* **17**, 7458-7464 (2009).
29. A. Kumar, C. Samaner, C. Cholsuk, T. Matthes, S. Pacal, Y. Oyun, A. Zand, R. J. Chapman, G. Saerens, R. Grange, S. Suwanna, S. Ates, T. Vogl, Polarization dynamics of solid-state quantum emitters. *ACS Nano* **18**, 5270-5281 (2024).
30. A. Gale, C. Li, Y. Chen, K. Watanabe, T. Taniguchi, I. Aharonovich, M. Toth, Site-specific fabrication of blue quantum emitters in hexagonal boron nitride. *ACS Photonics* **9**, 2170-2177 (2022).
31. S. Nedić, K. Yamamura, A. Gale, I. Aharonovich, M. Toth, Electron beam restructuring of quantum emitters in hexagonal boron nitride. *Adv. Opt. Mater.* **12**, 2400908 (2024).
32. A. Kumar, C. Cholsuk, A. Zand, M. N. Mishuk, T. Matthes, F. Eilenberger, S. Suwanna, T. Vogl, Localized creation of yellow single photon emitting carbon complexes in hexagonal boron nitride. *APL Mater.* **11**, 071108 (2023).

33. Q. Tao, J. Xie, J. Li, J. Zheng, P. Wang, K. Zhang, X. Zhao, T. Liang, Y. Gao, Y. Xie, N. Yao, C. Yu, F. Yu, L. Hu, N. Zhou, Continuous-wave laser induced formation of carbon-related defects in hexagonal boron nitride enhanced by surface plasmons and strain. *Adv. Funct. Mater.* **35**, 2423577 (2025).
34. H. Liu, N. Mendelson, I. H. Abidi, S. Li, Z. Liu, Y. Cai, K. Zhang, J. You, M. Tamtaji, H. Wong, Y. Ding, G. Chen, I. Aharonovich, Z. Luo, Rational control on quantum emitter formation in carbon-doped monolayer hexagonal boron nitride. *ACS Appl. Mater. Interfaces* **14**, 3189-3198 (2022).
35. A. Chatterjee, A. Biswas, A. S. Fuhr, T. Terlier, B. G. Sumpter, P. M. Ajayan, I. Aharonovich, S. Huang, Room-temperature high-purity single-photon emission from carbon-doped boron nitride thin films. *Sci. Adv.* **11**, eadv2899 (2025).
36. E. Blundo, A. Surrente, D. Spirito, G. Pettinari, T. Yildirim, C. A. Chavarin, L. Baldassarre, M. Felici, A. Polimeni, Vibrational properties in highly strained hexagonal boron nitride bubbles. *Nano Lett.* **22**, 1525-1533 (2022).
37. D. Wigger, R. Schmidt, O. Del Pozo-Zamudio, J. A. Preuß, P. Tonndorf, R. Schneider, P. Steeger, J. Kern, Y. Khodaei, J. Sperling, S. M. de Vasconcellos, R. Bratschitsch, T. Kuhn, Phonon-assisted emission and absorption of individual color centers in hexagonal boron nitride. *2D Mater.* **6**, 035006 (2019).
38. T. T. Tran, C. Elbadawi, D. Totonjian, C. J. Lobo, G. Grosso, H. Moon, D. R. Englund, M. J. Ford, I. Aharonovich, M. Toth, Robust multicolor single photon emission from point defects in hexagonal boron nitride. *ACS Nano* **10**, 7331-7338 (2016).
39. G. Jiang, X. Yuan, C. Liu, Y. Jiang, Q. Fu, C. Schimpf, T. Zheng, Q. Liu, D. Wan, Q. Zhang, F. Ding, D. Liu, Z. Ni, J. Lu, Planar optical antenna-driven brightness enhancement of interface-confined hexagonal boron nitride single-photon arrays for scalable room-temperature quantum chips. *ACS Nano* **19**, 33278-33287 (2025).
40. H. L. Stern, Q. Gu, J. Jarman, S. Eizagirre Barker, N. Mendelson, D. Chugh, S. Schott, H. H. Tan, H. Sirringhaus, I. Aharonovich, M. Atature, Room-temperature optically detected magnetic resonance of single defects in hexagonal boron nitride. *Nat. Commun.* **13**, 618 (2022).
41. J. Horder, S. J. U. White, A. Gale, C. Li, K. Watanabe, T. Taniguchi, M. Kianinia, I. Aharonovich, M. Toth, Coherence properties of electron-beam-activated emitters in hexagonal boron nitride under resonant excitation. *Phys. Rev. Appl.* **18**, 064021 (2022).
42. M. Onodera, K. Watanabe, M. Isayama, M. Arai, S. Masubuchi, R. Moriya, T. Taniguchi, T. Machida, Carbon-rich domain in hexagonal boron nitride: carrier mobility degradation and anomalous bending of the Landau fan diagram in adjacent graphene. *Nano Lett.* **19**, 7282-7286 (2019).
43. M. Neumann, X. Wei, L. Morales-Inostroza, S. Song, S. G. Lee, K. Watanabe, T. Taniguchi, S. Gotzinger, Y. H. Lee, Organic molecules as origin of visible-range single photon emission from hexagonal boron nitride and mica. *ACS Nano* **17**, 11679-11691 (2023).
44. N. Mendelson, D. Chugh, J. R. Reimers, T. S. Cheng, A. Gottscholl, H. Long, C. J. Mellor, A. Zettl, V. Dyakonov, P. H. Beton, S. V. Novikov, C. Jagadish, H. H. Tan, M. J. Ford, M. Toth, C. Bradac, I. Aharonovich, Identifying carbon as the source of visible single-photon emission from hexagonal boron nitride. *Nat. Mater.* **20**, 321-328 (2021).
45. S. S. Mohajerani, S. Chen, A. Alaei, T. Chou, N. Liu, Y. Ma, L. Xiao, S. S. Lee, E.-H. Yang, S. Strauf, Narrowband quantum light emission from oxygen-related color centers in hexagonal boron nitride. *ACS Photonics* **11**, 2359-2367 (2024).
46. F. Schaumburg, D. Plitt, T. Wagner, N. Wöhrl, M. Geller, G. Prinz, A. Lorke, Plasma-induced optically active defects in hexagonal boron nitride. *Appl. Phys. Lett.* **126**, 043103 (2025).

47. T. Petrenko, F. Neese, Analysis and prediction of absorption band shapes, fluorescence band shapes, resonance Raman intensities, and excitation profiles using the time-dependent theory of electronic spectroscopy. *J. Chem. Phys.* **127**, 164319 (2007).
48. T. Lu, F. Chen, Multiwfn: A multifunctional wavefunction analyzer. *J. Comput. Chem.* **33**, 580-592 (2012).
49. S. Hirata, M. Head-Gordon, Time-dependent density functional theory within the Tamm–Dancoff approximation. *Chem. Phys. Lett.* **314**, 291-299 (1999).
50. J. Zheng, A. V. Krasavin, R. Yang, Z. Wang, Y. Feng, L. Tang, L. Li, X. Guo, D. Dai, A. V. Zayats, L. Tong, P. Wang, Active control of excitonic strong coupling and electroluminescence in electrically driven plasmonic nanocavities. *Sci. Adv.* **11**, eadt9808 (2025).
51. J. Zheng, A. V. Krasavin, Z. Li, X. Guo, A. V. Zayats, L. Tong, P. Wang, Highly tunable optical response in dielectric-embedded plasmonic nanocavities. *Photon. Res.* **13**, 2593-2602 (2025).
52. S. Jin, F. Liu, I. Razdolski, T. W. Lo, Y. Wang, Z. Peng, K. Liang, Y. Zhu, W. Yao, A. V. Zayats, D. Lei, Plasmonic tuning of dark-exciton radiation dynamics and far-field emission directionality in monolayer $WSe_2$. *Sci. Adv.* **12**, eaea5781 (2026).
53. T. W. Lo, X. Chen, Z. Zhang, Q. Zhang, C. W. Leung, A. V. Zayats, D. Lei, Plasmonic nanocavity induced coupling and boost of dark excitons in monolayer $WSe_2$ at room temperature. *Nano Lett.* **22**, 1915-1921 (2022).
54. A. Boehmke Amoruso, R. A. Boto, E. Elliot, B. de Nijs, R. Esteban, T. Foldes, F. Aguilar-Galindo, E. Rosta, J. Aizpurua, J. J. Baumberg, Uncovering low-frequency vibrations in surface-enhanced Raman of organic molecules. *Nat. Commun.* **15**, 6733 (2024).
55. G. M. Akselrod, C. Argyropoulos, T. B. Hoang, C. Ciracì, C. Fang, J. Huang, D. R. Smith, M. H. Mikkelsen, Probing the mechanisms of large Purcell enhancement in plasmonic nanoantennas. *Nat. Photonics* **8**, 835-840 (2014).
56. N. Yamamoto, S. Ohtani, F. J. Garcia de Abajo, Gap and Mie plasmons in individual silver nanospheres near a silver surface. *Nano Lett.* **11**, 91-95 (2011).
57. T. Sannomiya, A. Konecna, T. Matsukata, Z. Thollar, T. Okamoto, F. J. Garcia de Abajo, N. Yamamoto, Cathodoluminescence phase extraction of the coupling between nanoparticles and surface plasmon polaritons. *Nano Lett.* **20**, 592-598 (2020).
58. G. Kresse, J. Furthmüller, Efficiency of ab-initio total energy calculations for metals and semiconductors using a plane-wave basis set. *Comput. Mater. Sci.* **6**, 15-50 (1996).
59. G. Kresse, D. Joubert, From ultrasoft pseudopotentials to the projector augmented-wave method. *Phys. Rev. B* **59**, 1758-1775 (1999).
60. J. P. Perdew, A. Ruzsinszky, J. Tao, V. N. Staroverov, G. E. Scuseria, G. I. Csonka, Prescription for the design and selection of density functional approximations: More constraint satisfaction with fewer fits. *J. Chem. Phys.* **123**, 062201 (2005).
61. J. P. Perdew, K. Burke, M. Ernzerhof, Generalized gradient approximation made simple. *Phys. Rev. Lett.* **77**, 3865-3868 (1996).
62. S. Grimme, S. Ehrlich, L. Goerigk, Effect of the damping function in dispersion corrected density functional theory. *J. Comput. Chem.* **32**, 1456-1465 (2011).
63. A. V. Krukau, O. A. Vydrov, A. F. Izmaylov, G. E. Scuseria, Influence of the exchange screening parameter on the performance of screened hybrid functionals. *J. Chem. Phys.* **125**, 224106 (2006).
64. C. Adamo, V. Barone, Toward reliable density functional methods without adjustable parameters: The PBE0 model. *J. Chem. Phys.* **110**, 6158-6170 (1999).
65. R. Krishnan, J. S. Binkley, R. Seeger, J. A. Pople, Self-consistent molecular orbital methods. XX. A basis set for correlated wave functions. *J. Chem. Phys.* **72**, 650-654 (1980).

66. F. Neese, F. Wennmohs, U. Becker, C. Riplinger, The ORCA quantum chemistry program package. *J. Chem. Phys.* **152**, 224108 (2020).

# Acknowledgements

The authors thank the Instrumentation and Service Center for Molecular Sciences at Westlake University for assistance with PL mapping and polarization spectroscopy measurements. We acknowledge Fang Chen and Jin Zhuang at the Zhejiang University Chemistry Instrumentation Center for SEM of the equipment supporting and analysis.
**Funding:** This work was supported by the National Natural Science Foundation of China (62005236, 62474156, 62405071), the Major Science and Technology Projects of Zhejiang Province (2025C01041), STI2030- Major Projects (2022ZD0212100); National Key Research and Development Program of China (2024YFB3613301); the Fundamental Research Funds for the Central Universities (226-2024-00142), and the Research Funds of Hangzhou Institute for Advanced Study, UCAS.

**Author contributions:** Conceptualization: Q.T., P.W., F.Y., and N.Z. Methodology: Q.T., T.L., S.Y., H.Z., Q.Z., Z.N., P.W., F.Y., and N.Z. Investigation: Q.T., F.Z., Z.L., Y.Y., S.L., Y.G., Z.W., and Y.L. Formal analysis: Q.T., F.Z., Q.Z., Z.N., F.Y., and N.Z. Visualization: Q.T., F.Z., Z.N., and N.Z. Funding acquisition: D.W., Q.Z., Z.N., and N.Z. Supervision: C.Y., Q.Z., Z.N., F.Y., L.H., and N.Z. Writing—original draft: Q.T. and N.Z. Writing—review & editing: Q.T., C.Y., Q.Z., Z.N., P.W., F.Y., L.H., and N.Z.

**Competing interests:** The authors declare that they have no competing interests.

**Data and materials availability:** All data needed to evaluate the conclusions in the paper are present in the paper and/or the Supplementary Information.

Supplementary Information for

# Electron-beam Writing of Spectrally Uniform Green Single-photon Emitters in Hexagonal Boron Nitride

## Authors

Qingsong Tao[1], Fuyi Zhou[2], Zhijie Li[3], Yihao Yan[3], Shuangyue Li[1], Yuelan Gao[1], Zijing Wu[1], Yizhou Liu[1], Tao Liang[1], Shuai Yuan[2], Dakun Wu[1], Hongzhi Zhou[1], Qi Zhang[5*], Zhenyi Ni[2*], Chunlei Yu[1,4], Pan Wang[6], Fei Yu[1,4*], Lili Hu[1,4], and Ning Zhou[1*]

## Affiliations

[1] School of Physics and Optoelectronic Engineering, Hangzhou Institute for Advanced Study, University of Chinese Academy of Sciences, Hangzhou 310024, China.

[2] State Key Laboratory of Silicon and Advanced Semiconductor Materials, School of Materials Science and Engineering, Zhejiang University, Hangzhou 310027, China.

[3] Anhui Province Key Laboratory of Scientific Instrument Development and Application, University of Science and Technology of China, Hefei 230026, China.

[4] Shanghai Institute of Optics and Fine Mechanics, Chinese Academy of Sciences, Shanghai 201800, China.

[5] Institute of Quantum Sensing, Institute of Fundamental and Transdisciplinary Research, School of Physics, Zhejiang Key Laboratory of R&D and Application of Cutting-Edge Scientific Instruments, State Key Laboratory of Ocean Sensing, Zhejiang University, Hangzhou 310027 China.

[6] New Cornerstone Science Laboratory, State Key Laboratory of Extreme Photonics and Instrumentation, College of Optical Science and Engineering, Zhejiang University, Hangzhou 310027, China.

[*]Corresponding author Email: qzquantum@zju.edu.cn; zyni@zju.edu.cn; yufei@siom.ac.cn; ningzhou@ucas.ac.cn.

**Table S1.**

| Methods | Spatial control | Representative ZPL distribution | ZPL wavelength selectivity | Thermal processing | Reported generation outcome | Ref. |
|---|---|---|---|---|---|---|
| MBE | Stochastic | 570-770 nm | Stochastic wavelengths | 1250℃ growth | Sparse emitters | 1 |
| MOVPE | Stochastic | 585 nm | Targeted wavelength | 1350℃ growth | High-density emitters | 1 |
| PLD | Stochastic | 560-590 nm | Stochastic wavelengths | 750℃ growth | Sparse emitters | 2 |
| Oxygen Environment Annealing | Stochastic | 540 nm | Targeted wavelength | 1000 ℃ annealing | Low-density green emitters | 3 |
| FIB | Site-selective | 500-700 nm | Stochastic wavelengths | 800-1150 ℃ annealing | Site-selective emitters | 4, 5 |
| Electron beam | Site-selective | 436 or 440 nm | Targeted wavelength | None | Site-selective emitters | 6 |
| Electron beam | Site-selective | 573-578 nm | Stochastic wavelengths | None | Site-selective emitters | 7, 8 |
| Strain | Site-selective | 624 nm | Targeted wavelength | 850 ℃ annealing | Single emitter generated from multiple strain sites | 9 |
| Strain | Site-selective | 530-610 nm | Stochastic wavelengths | None | Dense emitter generation along strained edges | 10 |
| Femtosecond Laser | Site-selective | 500-800 nm | Stochastic wavelengths | 900-1000 ℃ annealing | Site-selective, high-yield emitters | 11, 12 |
| Nanoindentation | Site-selective | 560-620 nm | Stochastic wavelengths | 850 ℃ annealing | Deterministic formation | 13 |
| CVD, C-doped Cu substrate | Stochastic | 560-650 nm | Stochastic wavelengths | 1090℃ growth | Sparse emitters | 14 |
| **Electron beam** | **Site-selective** | **536 nm** | **Targeted wavelength** | **None** | **Site-selective, high-yield green emitters** | ***This work*** |

**Table S1. Comparison of representative fabrication methods for SPEs in hBN.** The table summarizes representative hBN SPE fabrication methods in terms of spatial control, representative ZPL distribution, ZPL wavelength selectivity, thermal processing requirement, and reported generation outcome. The ZPL wavelength selectivity describes whether the emission wavelength can be selectively targeted or is broadly/stochastically distributed. The reported generation outcome is described qualitatively because yield metrics vary among studies, including emitter density, activated-site fraction, and large-area generation efficiency.

**Table S2.**

| Methods | ZPL (nm) | FWHM (nm) | Brightness | Lowest $g^{(2)}(0)$ | Ref. |
|---|---|---|---|---|---|
| Femtosecond laser | 500-800 | ~5 | $I_{sat}$ = 90 × $10^5$ cps<br>$P_{sat}$ = 764 μW | 0.09 | 12 |
| Femtosecond laser | 530-600 | ~6 | $I_{sat}$ = 86 × $10^5$ cps<br>$P_{sat}$ = 1.69 mW | 0.06 | 11 |
| Electron beam | 440 | ~11 | $I_{sat}$ = 3.6 × $10^5$ cps<br>$P_{sat}$ = 3.7 mW | 0.12 | 6 |
| Electron beam | 573-577 | ~15 | $I_{sat}$ = 0.47 × $10^5$ cps<br>$P_{sat}$ = 114 μW | 0.017 | 7, 15 |
| **Electron beam** | **536** | **~5** | **$I_{sat}$ = 22.4 × $10^5$ cps**<br>**$P_{sat}$ = 454 μW** | **0.08** | ***This work*** |
| $He^+$ beam (FIB) | 500-700 | ~4 | $I_{sat}$ = 3.3 × $10^5$ cps<br>$P_{sat}$ = 2.5 mW | 0.12 | 4 |
| $^{12}C^+$ implantation | 554-605 | ~9 | $I_{sat}$ = 2 × $10^5$ cps<br>$P_{sat}$ = 37 μW | 0.07 | 16 |
| Nanoindentation | 582-633 | ~13 | $I_{sat}$ = 11 × $10^5$ cps<br>$P_{sat}$ = 4.1 mW | 0.23 | 13 |

**Table S2. Comparison of room-temperature SPE performance in hBN created by representative fabrication methods.** The table summarizes representative room-temperature hBN SPEs in terms of ZPL wavelength range, linewidth, detected saturation brightness, and single-photon purity. Brightness values are reported as detected saturation count rates, $I_{sat}$, together with the corresponding saturation powers, $P_{sat}$, when available. The comparison highlights the combined performance of the EBW-written green emitters in this work, including the reproducible 536-nm ZPL, high detected brightness, and room-temperature antibunching.

**Fig. S1.**

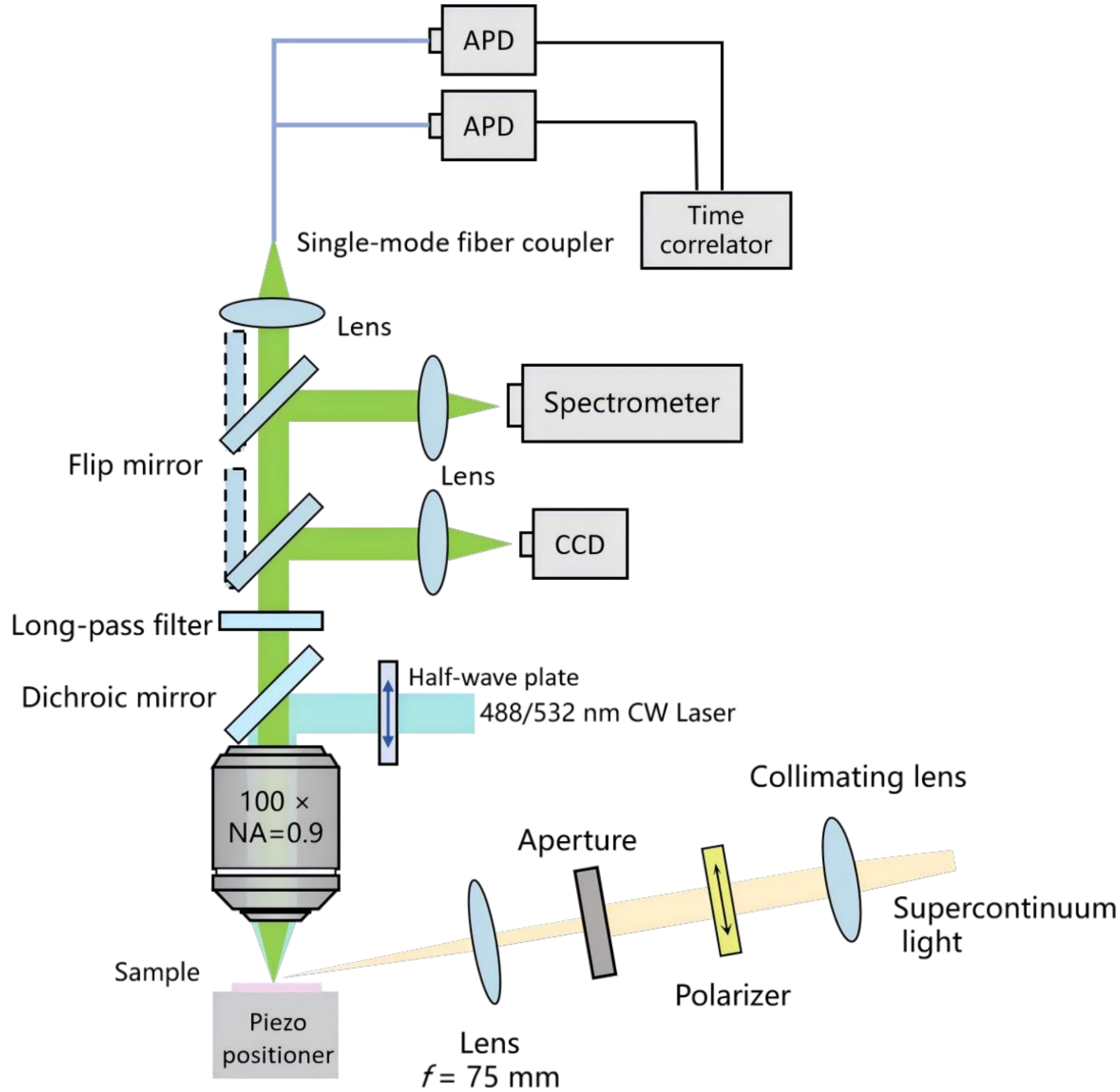

**Fig. S1. Schematic illustration of the customized confocal optical setup for optical characterization of EBW-written emitters.** Continuous-wave 488-nm and 532-nm lasers were used for excitation and focused onto the sample through a 100× objective. The collected signal was directed to different detection branches for wide-field imaging, PL/Raman spectroscopy, and Hanbury Brown and Twiss (HBT) autocorrelation measurements. Excitation-polarization-dependent measurements were performed by rotating the incident half-wave plate before the objective. Detailed optical components, filter sets, detector models, and acquisition conditions are described in the Methods section.

**Fig. S2.**

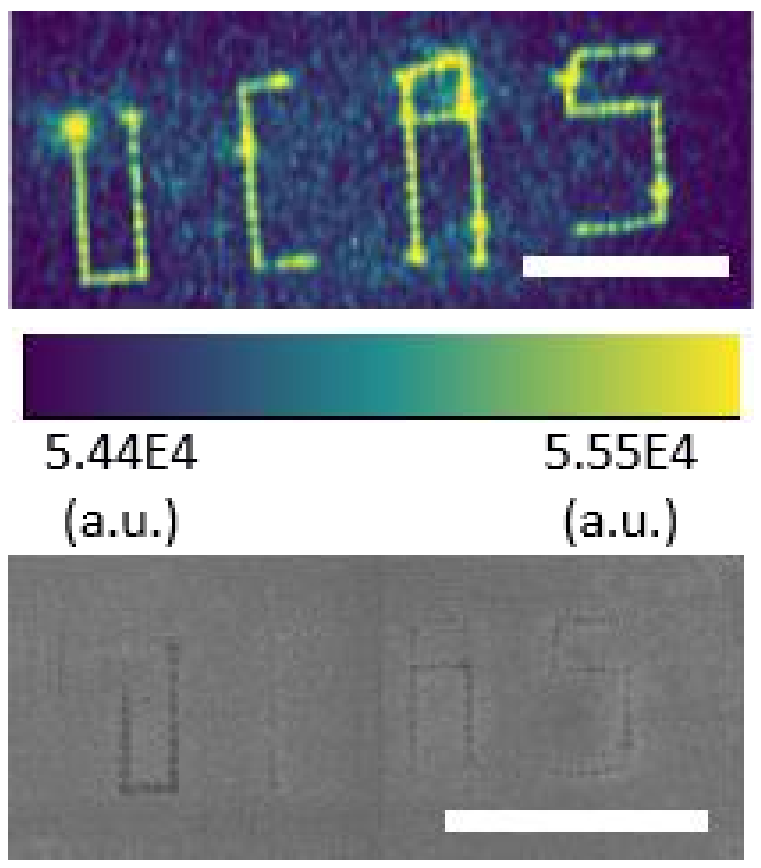

**Fig. S2. Spatial correlation of the optical emission and the morphology of EBW-defined patterns.** (Top) PL intensity map of the "UCAS" pattern generated by EBW in hBN. (Bottom) Corresponding SEM image of the same region. Scale bars: 20 μm.

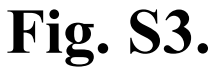

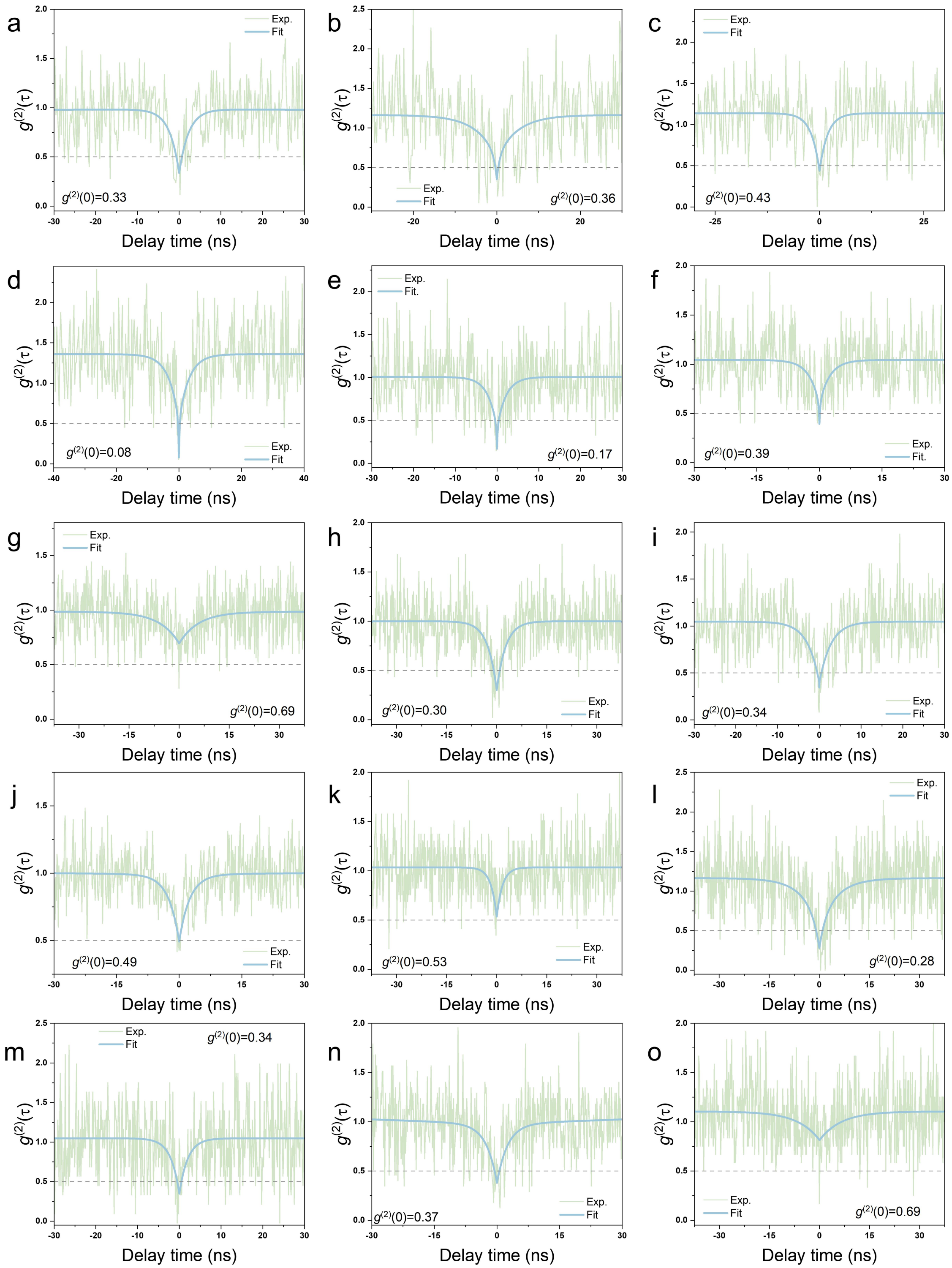

**Fig. S3. HBT characterization of multiple EBW-activated sites.** (a to o) Second-order autocorrelation function, $g^{(2)}(\tau)$, measured from distinct EBW-activated sites. The fitted $g^{(2)}(0)$ value is indicated in each panel. A large fraction of the measured spots exhibit a pronounced antibunching dip with $g^{(2)}(0) < 0.5$, confirming single-photon emission from those emitters.

**Fig. S4.**

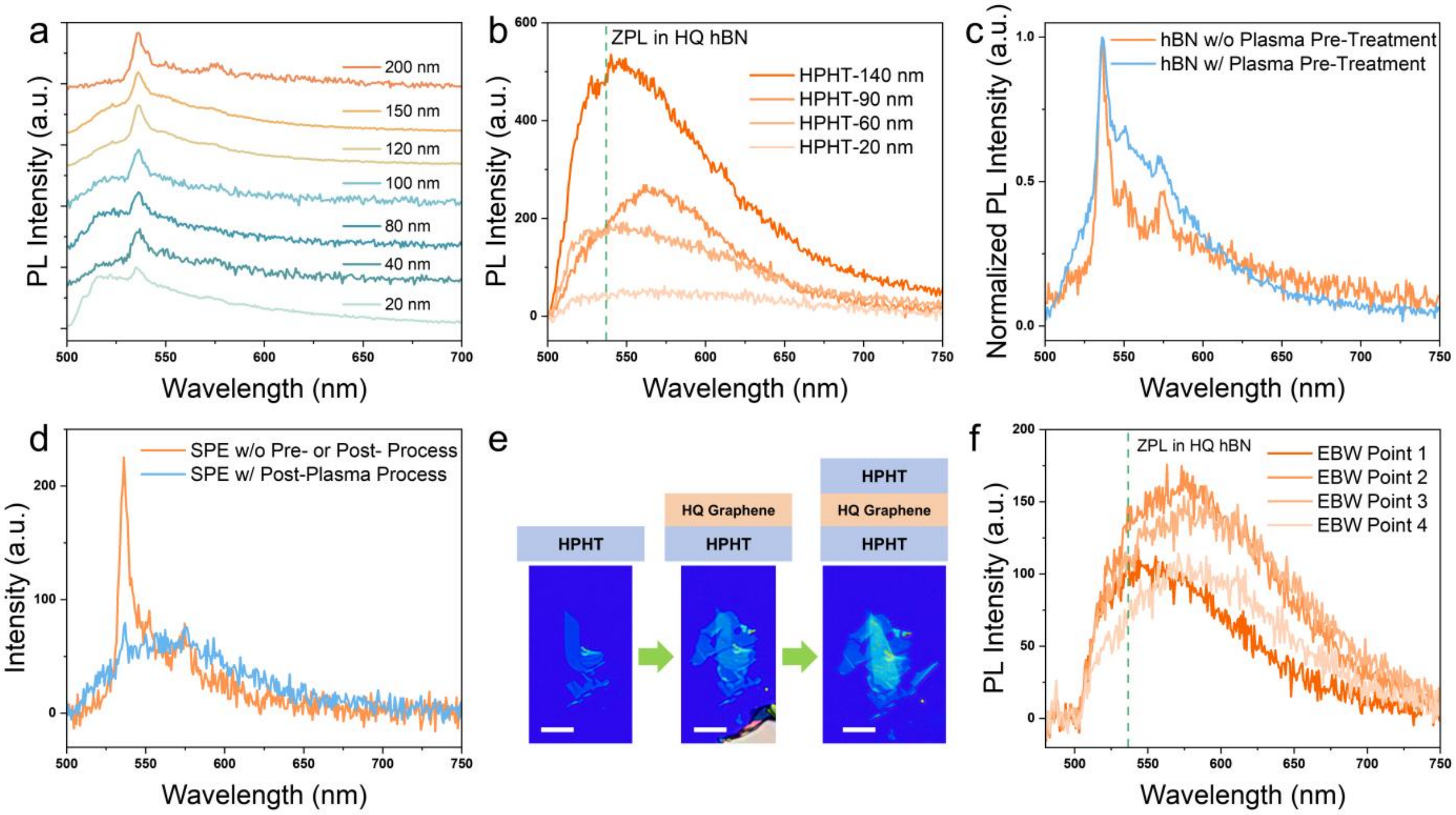

**Fig. S4. Thickness, material, and plasma-treatment effects on EBW activation of 536-nm emission.** (a) PL spectra recorded from EBW-irradiated HQ Graphene hBN flakes with thicknesses ranging from 20 to 200 nm. Together with Fig. 3a, these data show that the characteristic 536-nm emission can be reliably generated in flakes thicker than approximately 20 nm. (b) PL spectra of EBW-irradiated regions in HPHT hBN flakes with different thicknesses. Under the same EBW and optical characterization conditions, the characteristic 536-nm ZPL is absent, and only broad background emission is observed. (c) Normalized PL spectra showing that the characteristic 536-nm ZPL can still be generated in HQ Graphene hBN after an oxygen plasma pre-treatment. (d) PL spectra of an individual SPE before and after oxygen plasma post-treatment. The post-plasma process drastically reduces the emission intensity. (e) Schematic and microscope image illustrating the assembly of an HPHT-hBN/HQ Graphene hBN/HPHT-hBN sandwich structure. Scale bars, 10 µm. (f) PL spectra acquired from multiple EBW irradiation points on the sandwich structure shown in (e), showing that the target 536-nm SPE emission is not generated in this specific material configuration.

**Fig. S5.**

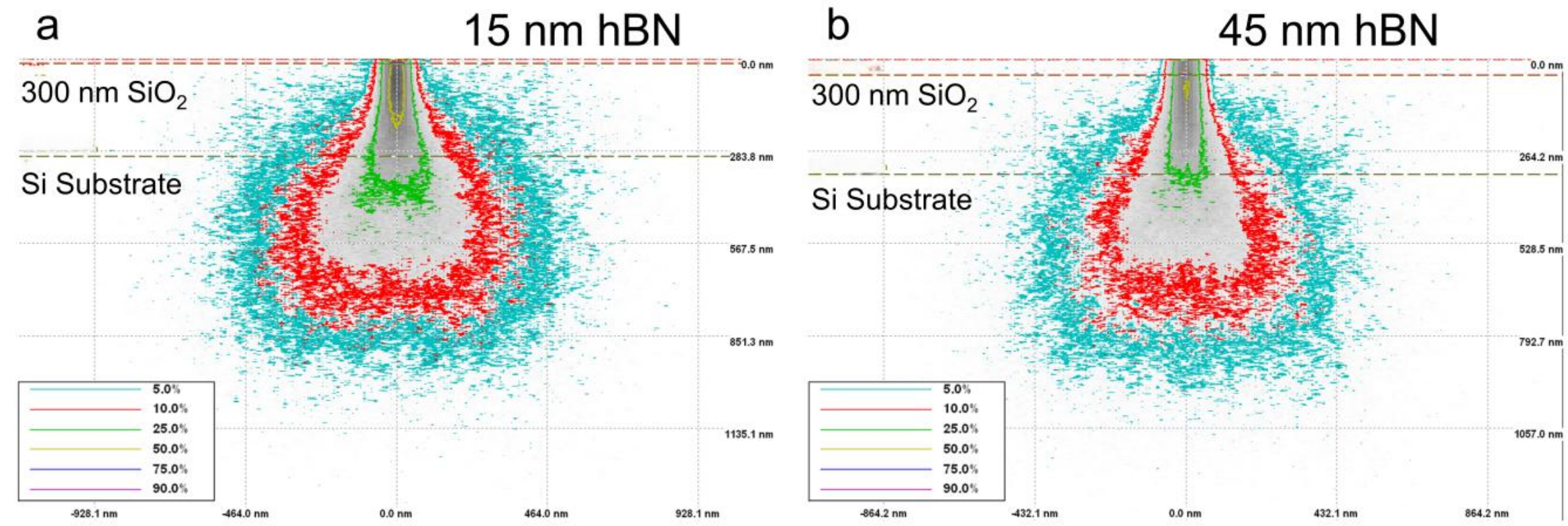

**Fig. S5. Monte Carlo simulations of the electron interaction volume within the hBN/$SiO_2$/Si structures.** Cross-sectional views of the simulated electron trajectories and energy dissipation profiles for incident electron beams interacting with an hBN flake of (a) 15 nm and (b) 45 nm thickness on $SiO_2$/Si substrate. Under the simulated conditions, the electron range exceeds the hBN thickness in both cases, and a substantial fraction of incident electrons reaches the underlying substrate. These simulations indicate that insufficient electron penetration is unlikely to be the sole reason for the absence of 536-nm emission in ultrathin hBN. Differences in local energy deposition, secondary-electron generation, charging, precursor volume, or defect-stabilization efficiency may also contribute to the observed thickness dependence.

Fig. S6.

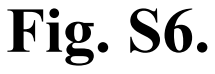

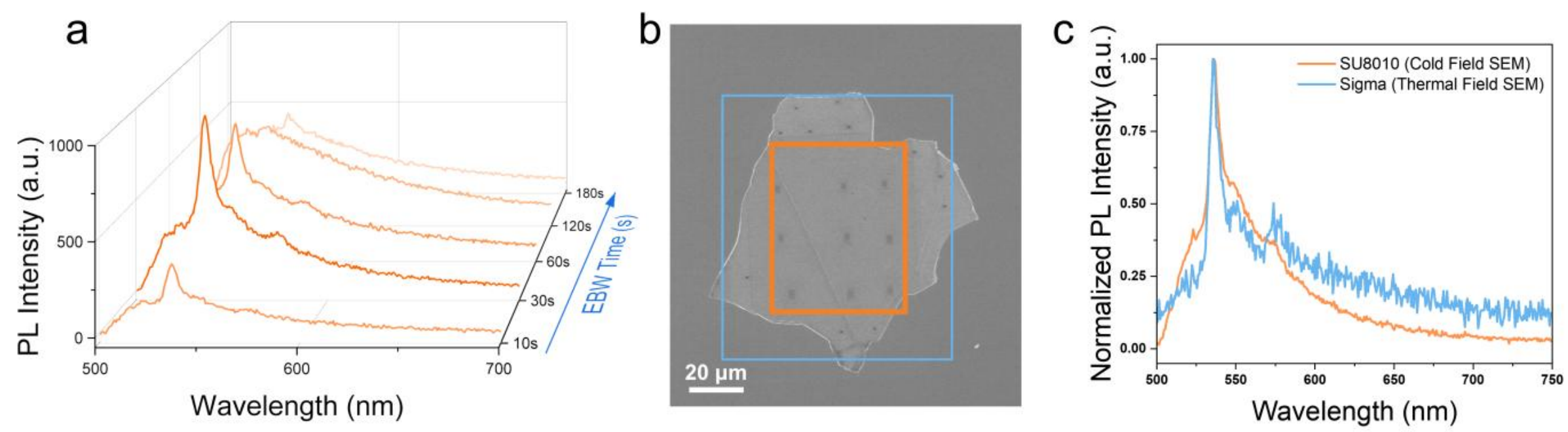

**Fig. S6. Irradiation-time dependence and SEM-system comparison of EBW activation in hBN.** (a) PL spectra acquired after EBW irradiation time from 10 s to 180 s. The 536-nm ZPL intensity reaches a maximum at ~30 s and decreases with prolonged irradiation, consistent with competition between defect activation and beam-induced damage, charge-state modification, or nonradiative pathway formation. (b) SEM image of an individual hBN flake patterned using two SEM systems. The point array inside the central blue rectangle was fabricated using a cold-field emission (CFE) SEM (Hitachi SU8010), whereas the peripheral array located between the orange and blue rectangles was irradiated using a thermal-field emission (TFE) SEM (Zeiss Sigma). (c) Normalized PL spectra showing qualitatively similar green emission generated using CFE-SEM and TFE-SEM systems.

**Fig. S7.**

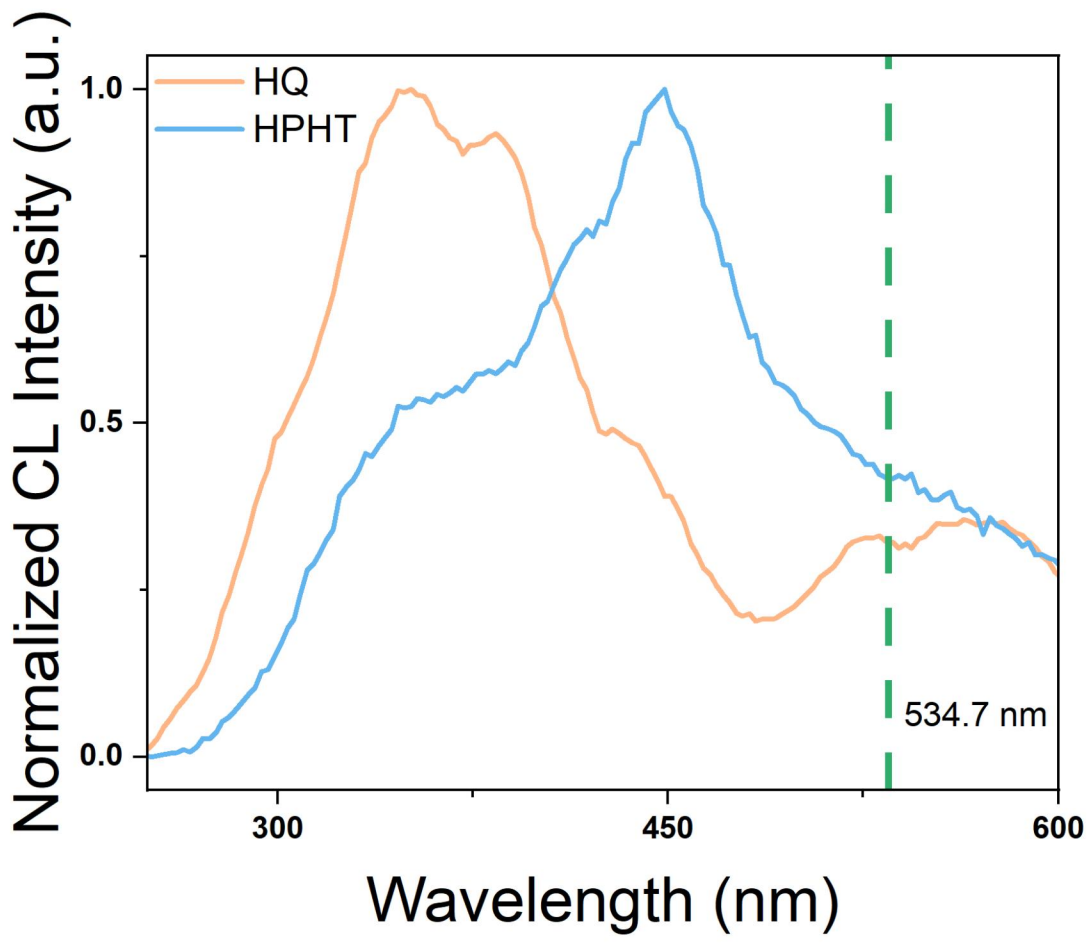

**Fig. S7. Room-temperature normalized CL spectra of thick HQ Graphene (orange curve) and HPHT (blue curve) hBN flakes.** CL spectra show that the HQ Graphene hBN exhibits a dominant broad emission band centered around 300 nm, a feature associated with carbon-related defect states in hBN[9]. A weak feature near 534.7 nm (green dashed line) is also observed and is spectrally close to the 536-nm PL ZPL of EBW-written emitters. In contrast, the tested HPHT hBN lacks these specific carbon-associated optical signatures. These data support the model in which carbon-related precursors contribute to EBW activation.

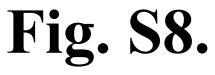

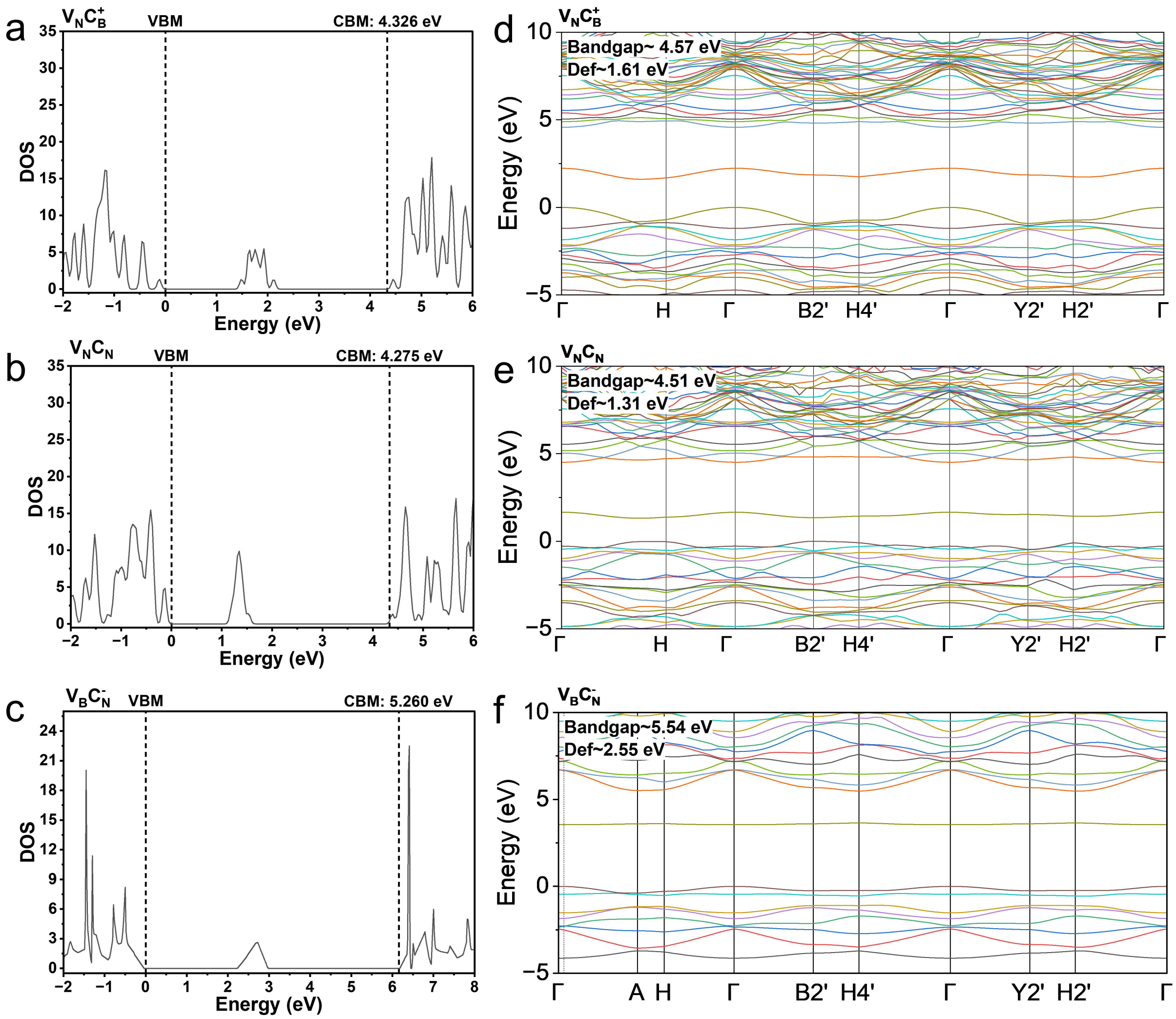

**Fig. S8. Electronic-structure comparison of candidate carbon-associated vacancy complexes in hBN.** Calculated density of states (a to c) and band structures (d to f) for $V_N C_B^+$, $V_N C_N$, and $V_B C_N^-$ defect complexes in hBN. These candidates were selected based on the experimentally observed carbon-related activation behavior and previous theoretical reports of carbon-associated hBN emitters with emission energies close to the 536-nm spectral region[7]. As an initial electronic-structure screening, the extracted defect-state energy separations for $V_N C_B^+$ and $V_N C_N$ are approximately 1.61 and 1.31 eV, respectively, both substantially smaller than the experimental ZPL energy of 2.31 eV. In contrast, $V_B C_N^-$ shows the closest transition-relevant defect-state separation, approximately 2.55 eV, motivating its selection for the vibrationally resolved TD-DFT emission-spectrum calculation shown in Fig. 4c.

**Fig. S9.**

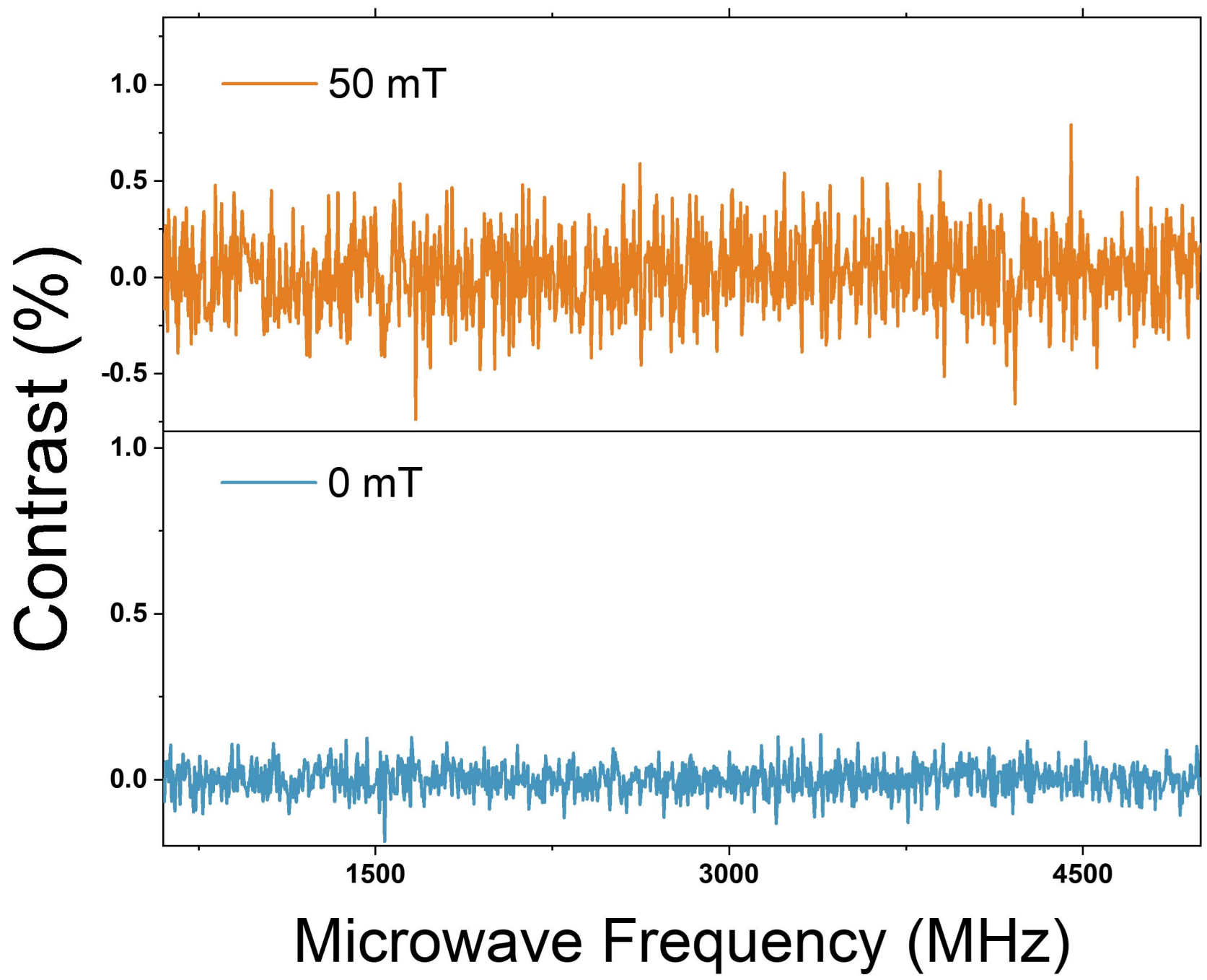

**Fig. S9. Optically detected magnetic resonance (ODMR) measurements of the 536-nm emitters.** ODMR spectra measured at zero magnetic field and under an applied 50 mT magnetic field. No resolvable magnetic-resonance contrast is observed within the explored frequency range of 600-5000 MHz under the present optical and microwave conditions. The absence of ODMR contrast does not determine the spin multiplicity of the defect and does not exclude spin transitions outside the explored detection window.

**References**

1. N. Mendelson, D. Chugh, J. R. Reimers, T. S. Cheng, A. Gottscholl, H. Long, C. J. Mellor, A. Zettl, V. Dyakonov, P. H. Beton, S. V. Novikov, C. Jagadish, H. H. Tan, M. J. Ford, M. Toth, C. Bradac, I. Aharonovich, Identifying carbon as the source of visible single-photon emission from hexagonal boron nitride. *Nat. Mater.* **20**, 321-328 (2021).
2. A. Chatterjee, A. Biswas, A. S. Fuhr, T. Terlier, B. G. Sumpter, P. M. Ajayan, I. Aharonovich, S. Huang, Room-temperature high-purity single-photon emission from carbon-doped boron nitride thin films. *Sci. Adv.* **11**, eadv2899 (2025).
3. H. Fartas, S. Hassani, J.-P. Hermier, N. D. Lai, S. Buil, A. Delteil, Reproducible generation of green-emitting color centers in hBN using oxygen annealing. *Appl. Phys. Lett.* **127**, 014001 (2025).
4. G. L. Liu, X. Y. Wu, P. T. Jing, Z. Cheng, D. Zhan, Y. Bao, J. X. Yan, H. Xu, L. G. Zhang, B. H. Li, K. W. Liu, L. Liu, D. Z. Shen, Single photon emitters in hexagonal boron nitride fabricated by focused helium ion beam. *Adv. Opt. Mater.* **12**, 2302083 (2024).
5. J. Ziegler, R. Klaiss, A. Blaikie, D. Miller, V. R. Horowitz, B. J. Aleman, Deterministic quantum emitter formation in hexagonal boron nitride via controlled edge creation. *Nano Lett.* **19**, 2121-2127 (2019).
6. C. Fournier, A. Plaud, S. Roux, A. Pierret, M. Rosticher, K. Watanabe, T. Taniguchi, S. Buil, X. Quélin, J. Barjon, J.-P. Hermier, A. Delteil, Position-controlled quantum emitters with reproducible emission wavelength in hexagonal boron nitride. *Nat. Commun.* **12**, 3779 (2021).
7. A. Kumar, C. Cholsuk, A. Zand, M. N. Mishuk, T. Matthes, F. Eilenberger, S. Suwanna, T. Vogl, Localized creation of yellow single photon emitting carbon complexes in hexagonal boron nitride. *APL Mater.* **11**, 071108 (2023).
8. A. Gale, C. Li, Y. Chen, K. Watanabe, T. Taniguchi, I. Aharonovich, M. Toth, Site-specific fabrication of blue quantum emitters in hexagonal boron nitride. *ACS Photonics* **9**, 2170-2177 (2022).
9. X. Chen, X. Yue, L. Zhang, X. Xu, F. Liu, M. Feng, Z. Hu, Y. Yan, J. Scheuer, X. Fu, Activated single photon emitters and enhanced deep-level emissions in hexagonal boron nitride strain crystal. *Adv. Funct. Mater.* **34**, 2306128 (2024).
10. N. V. Proscia, Z. Shotan, H. Jayakumar, P. Reddy, C. Cohen, M. Dollar, A. Alkauskas, M. Doherty, C. A. Meriles, V. M. Menon, Near-deterministic activation of room-temperature quantum emitters in hexagonal boron nitride. *Optica* **5**, 1128–1134 (2018).
11. L. Gan, D. Zhang, R. Zhang, Q. Zhang, H. Sun, Y. Li, C. Z. Ning, Large-scale, high-yield laser fabrication of bright and pure single-photon emitters at room temperature in hexagonal boron nitride. *ACS Nano* **16**, 14254-14261 (2022).
12. X. J. Wang, H. H. Fang, Z. Z. Li, D. Wang, H. B. Sun, Laser manufacturing of spatial resolution approaching quantum limit. *Light Sci. Appl.* **13**, 6 (2024).
13. M. Luo, J. Ge, P. Huang, Y. Yu, I. C. Seo, K. Lu, H. Sun, J. K. Tan, B. K. Tay, S. Kim, W. Gao, H. Li, D. Nam, Deterministic formation of carbon-functionalized quantum emitters in hexagonal boron nitride. *Nat. Commun.* **16**, 11450 (2025).
14. T. W. Tang, R. Ritika, M. Tamtaji, H. Liu, Y. Hu, Z. Liu, P. R. Galligan, M. Xu, J. Shen, J. Wang, J. You, Y. Li, G. Chen, I. Aharonovich, Z. Luo, Structured-defect engineering of hexagonal boron nitride for identified visible single-photon emitters. *ACS Nano* **19**, 8509-8519 (2025).

15. A. Kumar, C. Samaner, C. Cholsuk, T. Matthes, S. Pacal, Y. Oyun, A. Zand, R. J. Chapman, G. Saerens, R. Grange, S. Suwanna, S. Ates, T. Vogl, Polarization dynamics of solid-state quantum emitters. *ACS Nano* **18**, 5270-5281 (2024).
16. D. Zhong, S. Gao, M. Saccone, J. R. Greer, M. Bernardi, S. Nadj-Perge, A. Faraon, Carbon-related quantum emitter in hexagonal boron nitride with homogeneous energy and 3-fold polarization. *Nano Lett.* **24**, 1106-1113 (2024).